\documentstyle[manuscript,aps,epsfig,psfig]{revtex}
\begin{document}
\draft

\begin{center}
{\large {\bf Asymptotic and Exact Solutions of Perfect-Fluid Scalar-Tensor Cosmologies}}\vspace{0.5cm}

{A. Navarro$^1$, A. Serna$^{2,3}$ and J. M. Alimi$^3$}
\end{center}

{\small {\it $^1$Dept. F\'{\i}sica, Universidad de Murcia, E30071-Murcia, Spain\\
$^2$ F\'{\i}sica Aplicada, Universidad Miguel Hern\'andez, E03202-Elche, Spain\\
$^3$LAEC, Observatoire de Paris-Meudon (CNRS, URA-173), F92195-Meudon, France}}

\begin{abstract}
 We present a method which enables exact solutions to be found for flat 
homogeneous and isotropic scalar-tensor cosmologies with an arbitrary 
$\omega(\Phi)$ function and satisfying the general perfect fluid state 
equation $P=(\gamma-1)\rho c^2$. This method has been used to analyze 
a wide range of asymptotic analytical solutions at early and late 
times for different epochs in the cosmic history: false vacuum 
inflationary models, vacuum and radiation-dominated models, and 
matter-dominated models. We also describe the qualitative behavior of models
at intermediary times and give exact solutions at any time for some particular
scalar-tensor theories
\end{abstract}

\section{Introduction} \label{sec:intro}

The simplest generalization of Einstein's theory of gravity are 
scalar-tensor theories, which contain the metric tensor $g_{\mu\nu}$, 
and an additional scalar field $\phi$. The strength of the coupling 
between gravity and the scalar field is determined by an arbitrary 
coupling function $\omega(\phi)$. Although these theories have a long 
history \cite{Jordan-Fierz}, they have received a 
renewed interest both in cosmology and particle physics mainly because 
of three reasons. First, scalar-tensor theories are important 
in inflationary cosmology since they provide a way of exiting the 
inflationary epoch in a non-fine-tuned way \cite{Extended_Inflation}. 
Second, most of the recent attempts at unified models of fundamental 
interactions predict the existence of scalar partners to the tensor 
gravity of General Relativity (GR). This is the case of various 
Kaluza-Klein theories \cite{Kaluza-Klein}, as well as supersymmetric 
theories with extra dimensions and superstring theories 
\cite{Strings}, where scalar-tensor theories appear naturally as a 
low-energy limit. Finally, these theories can satisfy all the 
weak-field solar-system experiments and other present observations 
\cite{Will} to arbitrary accuracy, but they still diverge from GR in 
the strong limit. Thus, scalar-tensor theories provide an important 
framework for comparison with results of gravitational 
experiments which could take place in the near future (e.g., 
LIGOS \cite{LIGOS} or VIRGO \cite{VIRGO}).

 Analytical or numerical solutions for the scalar-tensor cosmological 
models are well known in the framework of some particular theories 
proposed in the literature. This is the case of {\em Brans-Dicke's} 
theory \cite{Brans-Dicke}, {\em Barker's} constant-$G$ theory 
\cite{Barker}, {\em Bekenstein's} variable rest mass theory 
\cite{Bekenstein}, or {\em Schmidt-Greiner-Heinz-Muller's} theory 
\cite{SGHM}. In the last years, a considerable effort has been devoted 
to investigate the cosmological models in more general scalar-tensor 
theories. For example, Burd and Coley \cite{Burd91} have used a 
dynamical system treatment to analyze the qualitative behavior of 
models where a constant (Brans-Dicke) coupling 
function is perturbed by a slight dependence on the scalar field.

A great improvement in the search of solutions of the scalar-tensor 
field equations has recently arisen in the form of methods which allow 
analytical integration through suitable changes of variables. This 
approach was first presented by Barrow \cite{Barrow93}, who generalized 
the method by Lorentz-Petzold \cite{LP} to solve the vacuum and 
radiation-dominated cosmological equations of any scalar-tensor theory. 
By using and extending this method, Serna \& Alimi \cite{SA96a} have 
performed an exhaustive study of the early-time behavior during the 
radiation-dominated epoch of scalar-tensor theories where 
$\omega(\phi)$ is a monotonic, but arbitrary, function of $\phi$.

A further step in the search of exact scalar-tensor cosmological 
solutions was provided by Barrow, Mimoso and others \cite{Barrow94} 
who developed a method to solve the Friedmann-Robertson-Walker 
field equations in models with a perfect fluid satisfying the equation of 
state $P=(\gamma-1)\rho c^2$ (with $\gamma$ a constant and 
$0\leq\gamma\leq2$). In this method, solutions are found through a 
generating function $g$ which is indirectly related to the coupling 
function $\omega(\phi)$. A systematic study of the asymptotic behavior 
of scalar-tensor theories, although possible, is then difficult to be 
performed since the connection between $g$ and $\omega(\phi)$ is only 
known after completely solving the field equations. Exact solutions 
during all the epochs of cosmological interest have been nevertheless 
derived for some classes of scalar-tensor theories \cite{Barrow94,BP96}. 

We will present in this paper an alternative method, also based on a 
suitable change of variables, which enables exact solutions to be found 
for homogeneous and isotropic scalar-tensor cosmologies with an 
arbitrary $\omega(\phi)$ function. Unlike the previous methods (except for
some studies of the late time behavior of scalar-tensor theories 
\cite{Comer}), the generating function will be now the time dependence of 
the coupling function itself, what will allow for an easier analysis of 
the early and late time asymptotic solutions during all the main 
epochs in the universe evolution.

The paper is arranged as follows. We begin outlining the scalar-tensor
theories (Sect. \ref{sec:ST}) and we then present a method to build up 
homogeneous and isotropic cosmological models in their framework 
(Sect. \ref{sec:method}). This method is then applied to analyze the early 
(Sects. \ref{asymptotic} and \ref{asymptotic1}) and late time 
solutions (Sect. \ref{asymptotic2}) for a wide range of theories. Finally, 
conclusions and a summary of our results are given in Sec. 
\ref{sec:conclus}.

\section{Scalar-Tensor Cosmologies} \label{sec:ST}

\subsection{Field equations}
The most general action describing a massless scalar-tensor theory of
gravitation is \cite{BWN}

\begin{equation}
S = \frac{1}{16\pi}\int (\phi{\cal R} -\frac{\omega(\phi)}{\phi}
\phi_{,\mu} \phi^{,\mu}) \sqrt{g} d^{4}x + S_{M}, \label{eq:SJ}
\end{equation}

\noindent 
where ${\cal R}$ is the curvature scalar of the metrics $g_{\mu\nu}$,
$g\equiv det(g_{\mu\nu})$, and $\phi$ is a scalar field. The strength of the coupling between gravity and the 
scalar field is determined by the arbitrary function $\omega(\phi)$, usually termed the {\it coupling function}.
Each specific choice of the $\omega(\phi)$ function defines a particular scalar-tensor theory. The simplest case 
is Brans-Dicke's theory, where $\omega$ is a constant.

The variation of Eq.\ (\ref{eq:SJ}) with respect to $g_{\mu\nu}$
and $\phi$ leads to the field equations:

\begin{mathletters}
\label{eq:field}
\begin{eqnarray}
 {\cal R}_{\mu\nu}-\frac{1}{2}g_{\mu\nu} {\cal R} &=& -
 \frac{8\pi}{\phi} T_{\mu\nu} - \frac{\omega}{\phi^{2}}(\phi_{,\mu}
 \phi_{,\nu} - \frac{1}{2} g_{\mu\nu} \phi_{,\kappa}
 \phi^{,\kappa})\nonumber\\ & & -\frac{1}{\phi} (\phi_{,\mu;\nu} -
 g_{\mu\nu} \Box \phi )
\label{eq:field1} \\
(3+2\omega)\Box\phi &=& 8\pi T - \omega'
\phi_{,\kappa}\phi^{,\kappa} \label{eq:field2}
\end{eqnarray}
\end{mathletters}

\noindent
which satisfy the usual conservation law

\begin{equation}
T^{\mu\nu}_{;\nu} = 0, 
\end{equation}

\noindent
where $T^{\mu\nu}$ is the energy-momentum tensor, $\omega'$ denotes
$d\omega/d\phi$ and $\Box\phi\equiv g^{\mu\nu}\phi_{,\mu;\nu}$.

\subsection{Cosmological models}
In order to build up cosmological models in the framework of scalar-tensor gravity theories, we consider a 
homogeneous and isotropic universe. The line-element has then a Robertson-Walker form:

\begin{equation}
 ds^{2} = -c^{2} dt^{2} + a^{2}(t)\left[\frac{dr^{2}}{1 - Kr^{2}} +
 r^{2}d\Omega^{2}\right]
\end{equation}

\noindent 
and the energy-momentum tensor corresponds to that of a perfect fluid

\begin{equation}
T^{\mu\nu} = (\rho+P/c^{2})u_{\mu}u_{\nu} + Pg_{\mu\nu}\label{Tmunu}
\end{equation}
where $K=0, \pm 1, a(t)$ is the scale factor, $\rho$ and $P$ are the
energy-mass density and pressure, respectively, and $u_{\mu}$ is the
4-velocity of the fluid. 

By writing the equation of state as 

\begin{equation}
P=(\gamma-1)\rho c^2 \hspace{1cm}(0\leq\gamma\le 2)
\label{eqstate}
\end{equation}
and assuming a flat Universe ($K=0$), the field equations (\ref{eq:field}) become

\begin{mathletters}\label{eq:basic}
\begin{eqnarray}
\frac{8\pi G\rho }{3\Phi} +\left(\frac{\dot{\Phi}}{2\Phi}\right)^2 
\frac{(3+2\omega)}{3}=\left[\frac{\dot{a}}{a} +
\frac{\dot{\Phi}}{2\Phi}\right]^2 \label{eq:basic1}\\
\ddot{\Phi}+\left[3\frac{\dot{a}}{a}+\frac{\dot{\omega}}{3+2\omega}\right] \dot{\Phi} =
\frac{8\pi G\rho(4-3\gamma)}{(3+2\omega)} \label{eq:basic2}
\end{eqnarray}
\end{mathletters}

\noindent
where $G$ is Newton's gravitational constant, $\Phi\equiv G \phi$, and dots mean time derivatives. In addition, 
the conservation 
equation (\ref{Tmunu}) implies
\begin{equation}
\rho\propto a^{-3\gamma}\label{eq:conserv}
\end{equation}

\section{Integration method}\label{sec:method}

The main difficulty for finding complete or asymptotic solutions of the 
scalar-tensor cosmological equations is the presence of a source term in 
the right hand side of the wave equation (\ref{eq:basic2}). The only 
case in which these equations can be solved for universes of all 
curvatures is $\gamma=4/3$ (vacuum and radiation-dominated models), 
where Eq. (\ref{eq:basic2}) becomes sourceless \cite{Barrow93}. For 
general perfect fluids ($\gamma\neq4/3$), Eqs. (\ref{eq:basic}) can be 
solved only for flat universes and using 'indirect' methods \cite{Barrow94}
where a particular scalar-tensor theory is defined through some generating 
function, $g$, instead of the coupling function $\omega(\Phi)$. 

We will now present a 'semi-indirect' method where the coupling function $\omega$ is used as the generating 
function defining each particular theory. However, unlike the 'direct' method available for radiation-dominated 
($\gamma=4/3$) epochs, the form of $\omega(\Phi)$ as a function of $\Phi$ is not initially known, but only its 
dependence on a 'time' variable $\tau$.

\subsection{Integration method}
In order to solve Eqs. (\ref{eq:basic}), we introduce a new time variable $\tau$

\begin{equation}
d\tau=\sqrt{\Gamma} a^{3(1-\gamma)}dt\label{tau}
\end{equation}

\noindent
and two dynamical variables

\begin{eqnarray}
y &\equiv& a^{3(2-\gamma)} \Phi'\sqrt{\mid\!3+2\omega\!\mid/3}\label{y}\\
x &\equiv& a^{3(2-\gamma)}\Phi\label{x}
\end{eqnarray}

\noindent
where primes denote differentiation with respect to $\tau$ and $\Gamma\equiv 8\pi G\rho a^{3\gamma}/3=$ 
constant.

Using these variables, the field equations (\ref{eq:basic}) reduce to
\begin{equation}
y' = (4-3\gamma) \frac{\sigma}{\sqrt{\mid\!3+2\omega\!\mid/3}}
\label{eq:y'}
\end{equation}
and
\begin{equation}
x+\sigma y^2/4=\frac{1}{(6-3\gamma)^2}
\left[ x'+\sigma \frac{(y^2)'}{4}\right]^2
\label{eq:x'}
\end{equation}
where $\sigma\equiv\mbox{sign}(3+2\omega)$.

By introducing the function

\begin{equation}
f^2\equiv x+ \sigma y^2/4, \label{f}
\end{equation}
Eq. (\ref{eq:x'}) can be written in a more compact way as
\begin{equation}
 (f^2)'= 3(2-\gamma) f \label{eq:f'}
\end{equation}

Integration of Eqs. (\ref{eq:y'}) and (\ref{eq:f'}) is then straightforward. The solutions are

\begin{eqnarray}
y &=& \sigma(4-3\gamma)\int_{0}^{\tau}
\frac{d\tau}{\sqrt{\mid\!3+2\omega\!\mid/3}}+y_0 
\label{eq:y}\\
f &=& \frac{3}{2}(2-\gamma)\tau+f_0 \label{eq:f}
\end{eqnarray}
\noindent
where $y_0$ and $f_0$ are integration constants. We see from Eqs. 
(\ref{eq:y}) and (\ref{eq:f}) that the particular cases $4-3\gamma=0$ 
and $3(2-\gamma)=0$ imply $y=$constant and $f=$constant, respectively. 
In the first case, complete exact solutions can be easily found from Eq. 
(\ref{y}) because the second integration implied by Eq. (\ref{eq:y}) is not 
needed. Since the $3+2\omega$ function, as well as the constant factors $(4-3\gamma)$ and $3(2-\gamma)$ will 
appear several times throughout this paper, we will denote

\begin{eqnarray}
W(\tau) &\equiv& (3+2\omega)/3\\
\gamma_1 &\equiv& 3(2-\gamma)\\
\gamma_2 &\equiv& 4-3\gamma
\end{eqnarray}

\noindent
Constants $\gamma_1$ and $\gamma_2$ are known at any epoch in the universe evolution. A hypothetical false-vacuum era would 
correspond to $\gamma=0$ ($\gamma_1=6$, $\gamma_2=4$), while the vacuum ($\Gamma=0$) and radiation-
dominated ($\Gamma>0$) eras correspond to $\gamma=4/3$ ($\gamma_1=2$, $\gamma_2=0$). Finally, the 
matter-dominated era is defined by $\gamma=1$ ($\gamma_1=3$, $\gamma_2=1$). 

We will see now how Eqs. (\ref{tau})-(\ref{eq:f}) enable exact solutions to be found for the scalar-tensor field 
equations (\ref{eq:basic}). To that end, we will consider separately the cases $\gamma<4/3$ ($\gamma_2>0$) 
and $\gamma=4/3$ ($\gamma_2=0$).

\subsubsection{Case $\gamma<4/3$ (general perfect fluids)}

When $\gamma_2>0$, we can use Eqs. (\ref{y})--(\ref{eq:y'}) to write
\begin{equation}
\frac{\Phi'}{\Phi}=\frac{\sigma}{\gamma_2}\frac{yy'}{x}
\end{equation}

By integrating this equation, and using Eq. (\ref{f}), we find
\begin{equation}
\Phi=\Phi_1 (e^{J}/\mid\!x\!\mid)^{2/\gamma_2} \label{Phi(tau)}
\end{equation}

\noindent
where sign$(\Phi)=$ sign$(x)$ (see Eq. \ref{x}), and
\begin{equation}
J\equiv \gamma_1 \int_{\tau_1}^{\tau}\frac{f}{x}d\tau\label{J}
\end{equation}

\noindent
with $\Phi_1$ and $\tau_1$ being integration constants.

From Eqs. (\ref{x}) and (\ref{Phi(tau)}), the scale factor is then given by
\begin{equation}
a = a_1 (\mid\!x\!\mid /e^{2J/\gamma_1})^{1/\gamma_2} \label{a(tau)}
\end{equation}
where $a_1>0$ is another integration constant.

The $a(t)$, $\Phi(t)$ and $\omega(\Phi)$ functions can be then obtained 
in the following way: {\it i}) we start by choosing a particular scalar-tensor 
theory, that is, by specifying $3+2\omega$ as a function of $\tau$; {\it ii}) 
we compute $y(\tau)$ from Eq. (\ref{eq:y}); {\it iii}) $x(\tau)$ is then 
obtained from Eqs. (\ref{f}) and (\ref{eq:f}); {\it iv}) $\Phi(\tau)$ from 
Eq. (\ref{Phi(tau)}); {\it v}) $\omega(\Phi)$ by inverting $\Phi(\tau)$ and 
using $3+2\omega(\tau)$; {\it vi}) $a(\tau)$ from Eq. (\ref{a(tau)}); and 
{\it vii}) $\tau(t)$ from Eq. (\ref{tau}). Obviously, we assume that the 
corresponding expressions are all integrable and invertible.

Once $a(\tau)$ is known, we can find other functions of cosmological interest.
For example, from Eq. (\ref{tau}), the Hubble parameter $H\equiv\dot{a}/a$ 
is given by

\begin{equation}
H=\sqrt{\Gamma}a^{3(1-\gamma)}\left(\frac{a'}{a}\right) .\label{H}
\end{equation}

Differentiating Eqs. (\ref{x}) and (\ref{f}), and using Eqs. (\ref{y}) and 
(\ref{eq:f'}) to eliminate $\Phi$ and $f$ derivatives, $h\equiv a'/a$ can be written in terms of $f(\tau)$, 
$y(\tau)$, and $\omega(\tau)$ as

\begin{equation}
h=\frac{1}{x}\left[ f-\frac{y}
{2\sqrt{\mid\!3+2\omega\!\mid/3}}\right]\label{h}
\end{equation}

Another function of interest is the 
speed-up factor $\xi\equiv H/H_{GR}$, which compares the expansion rate, $H$, to that obtained in the framework of GR at the same temperature, $H_{GR}$. To obtain this quantity, we use Eq. (\ref{H}) and 
$H_{GR}^{2}=8\pi G\rho/3=\Gamma a^{-3\gamma}$, so that

\begin{equation}
\xi= a^{\gamma_2/2}a'= a^{\gamma_1/2}h\label{xi}
\end{equation}

\subsubsection{Case $\gamma=4/3$ (vacuum and radiation-dominated models)} \label{4a}

When $\gamma_2=0$, Eqs. (\ref{h})-(\ref{xi}) hold but 
now with a constant $y$ value (see Eq. \ref{eq:y'}): 
\begin{equation}
y=y_0
\end{equation}
Eqs. (\ref{y}) and (\ref{f}) then become
\begin{eqnarray}
\Phi'a^2&=& \sqrt{3}y_0 \mid\!3+2\omega\!\mid^{-1/2}\label{Phirad}\\
x&=&(\tau+f_0)^2-\sigma y_{0}^{2}/4\label{xrad}
\end{eqnarray}
which coincide with Eqs. (15)-(16) of Serna \& Alimi \cite{SA96a} 
(see also ref. \cite{Barrow93}). 

Using Eq. (\ref{x}) to eliminate the scale factor in Eq. (\ref{Phirad}), and integrating Eqs. (\ref{Phirad})-
(\ref{xrad}), we obtain (for $\Gamma\neq0$):

\begin{eqnarray}
\int\mid\!3&+&2\omega\!\mid^{1/2}\frac{d\Phi}{\Phi}= \\
&=&\left\{
\begin{array}{ll}
\sqrt{3}\ln\left(\frac{\tau+f_0-A}{\tau+f_0+A}\right) &(\sigma>0)\\
2\sqrt{3}\left[\mbox{atan}\left(\frac{\tau+f_0}{A}\right)-
\mbox{sign}(A)\pi/2\right] & (\sigma<0)
\end{array}\right. \nonumber 
\end{eqnarray}

\noindent 
where $A\equiv y_0/2$.

From a similar procedure (but using $a d\tau\equiv dt$), it is straightforward to show that, when $\Gamma=0$ and $\sigma>0$

\begin{equation}
\int\mid\!3+2\omega\!\mid^{1/2}\frac{d\Phi}{\Phi}=
\sqrt{3}\mbox{ sign}(A)\ln[|A|(\tau+f_0)]
\end{equation}

\noindent
while no solutions exist for $\sigma<0$.

Since a detailed study of the radiation-dominated asymptotic solutions has 
been performed in a previous paper \cite{SA96a}, we will give 
here just a tabulated summary of results (see Table 1). 

\subsection{Setting of constants}\label{origin}

Since the origin of times can be chosen according to any arbitrary 
criterion, an analysis of asymptotic solutions at early times ($\tau 
\rightarrow0$) would be meaningless if we do not clearly specify what 
means $\tau=0$ for us. In this paper, we will set the origin of times by using a criterion similar to 
that of Serna \& Alimi \cite{SA96a}. Using 
the notation of the present paper and assuming that the sign of $3+2\omega$ never changes during the Universe evolution, such a criterion can be expressed as follows:

\subsubsection{Models with $3+2\omega>0$ ($\sigma=+1$)}\label{origine1}

As we will see now, these models are characterized by the fact that $x$ vanishes at least at one finite value of $\tau$. Consequently, we 
take the origin of times so that the most recent zero in $x$ occurs at 
$\tau=0$ ($x(0)=0$, and $x>0$ for any $\tau>0$). We see from Eq. 
(\ref{x}) that such a criterion ensures two important features of our 
solutions: 1) a singular point either in the scale factor $a$ or in the scalar 
field $\Phi$ defines the origin of times, 2) both $a$ and $\Phi$ have 
positive values (strictly) for any $\tau>0$.

From Eq. (\ref{f}), we see that condition $x(0)=0$ can be written as
\begin{equation}
f_{0}^{2}=\sigma y_{0}^{2}/4\; . \label{generalcondit}
\end{equation}

Throughout this paper, the subindex $0$ denotes value at $\tau = 0$. For $\sigma=+1$ and any value 
of $y_0$ (positive, negative or vanishing), we then find that the $f_0$ 
value can always be chosen so that $x(0)=0$. 

The condition 
$x>0$ for any $\tau>0$ implies that $f_0\geq 0$. As a matter of 
fact, if $f_0$ were negative, the function $f=(\gamma_1/2)\tau+f_0$ 
would vanish at a positive time $\tau_{\ast}=-2f_0/\gamma_1$. Thus, 
$x=f^{2}-y^2/4=-y^2/4<0$ at $\tau_{\ast}>0$, in contradiction with the 
requirement $x>0$ for any $\tau>0$. By combining $f_0\geq0$ with the 
condition (\ref{generalcondit}), we then find that integration constants 
must be chosen as

\begin{equation}
f_0=|y_0|/2 \hspace{0.5cm}(\sigma=+1)\; .
\label{f_0}
\end{equation}

Another important consequence of the requirement $x>0$ for any $\tau>0$ is that $f'_0\ge |y|'_0/2$. As a matter of fact, when this requirement is combined with Eq. (\ref{f}), one obtains that $f(\tau)>|y(\tau)|/2$. Since $f_0=|y_0|/2$, the last inequality can 
only be satisfied if $f'_0\ge |y|'_0/2$. Consequently, from Eqs. (\ref{eq:y'})
and (\ref{eq:f}), we find that

\begin{equation}
\sqrt{|W_0|}\ge\left\{\begin{array}{ll}
\gamma_2/\gamma_1 & \mbox{(if $y_0\ge 0$)}\\
0 & \mbox{(if $y_0<0$)}
\end{array}\right.\hspace{0.5cm}(\sigma=+1)\; .
\label{W_0}
\end{equation}

We note however that Eqs. (\ref{f_0}) and (\ref{W_0}) are just necessary, 
but not 
sufficient, conditions to satisfy our criterion for the origin of times. 

\subsubsection{Models with $3+2\omega<0$ ($\sigma=-1$)}\label{origine2}

When $3+2\omega<0$ (i.e., $\sigma=-1$), the only choice satisfying 
Eq. (\ref{generalcondit}) is $f_0=y_0=0$. That is,
\begin{equation}
f_0=0\hspace{0.5cm}\mbox{(if $y_0=0$ and $\sigma=-1$)}\; .\label{f0s<0}
\end{equation}

For $\sigma=-1$ and non-vanishing (positive or negative) $y_0$ values, the condition (\ref{generalcondit}) is not 
satisfied and $x$ then never vanishes. In that case, we will see now that the scale factor has a 
non-zero minimum value and models are all non-singular. Consequently, 
we will take the origin of times so that the last stationary point of $a$ 
occurs at $\tau=0$.

The existence of maxima or minima in the time evolution of the scale 
factor has been previously analyzed only for vacuum and 
radiation-dominated scalar-tensor cosmologies \cite{BP96}. The 
integration method presented in this paper also allows for an easy study 
of the existence of stationary points at any epoch of the universe 
evolution.

The condition for the scale factor to contain a stationary point, $a'=0$ 
with $a>0$, is equivalent to $h_0=0$, which leads to
\begin{equation}
f_0=\frac{y_0}{2\sqrt{|W_0|}}
\hspace{0.5cm}\mbox{(if $y_0\ne0$ and $\sigma=-1$)}\; \label{minimum}
\end{equation}

\noindent
as the appropriate condition to take the origin of times so that $a$ has 
initially a stationary point. 

We must note that the integration constants $f_0$ and $y_0$ must have 
finite values. Otherwise, solutions would diverge at any time. Consequently, 
Eq. (\ref{minimum}) implies that the possibility $W_0
\rightarrow0$ is excluded when $y_0\neq0$ and $3+2\omega<0$.

\subsection{The coupling function}\label{Coupling}

\subsubsection{Representation for early and late times}

We will use the same representation for the coupling function 
as in our previous work \cite{SA96a,SA96b} concerning 
radiation-dominated cosmological models. That is, 

\begin{equation}
\mid\!3+2\omega\!\mid = 
(3/\lambda^{2})(k+\mid\!\Phi-1\!\mid^{-\delta}) \label{eq:w32}
\end{equation}

\noindent
where $1/2<\delta<2$ (to allow for convergence towards GR), 
$\lambda>0$, and $k$ are arbitrary constants. This form gives an exact 
description for most of the particular scalar-tensor theories proposed 
in the literature and, in addition, it contains all the possible early 
behaviors of any theory where $\omega(\Phi)$ is a monotonic, but 
arbitrary, function of $\Phi$ (see Appendix A).

Barrow \& Parsons \cite{BP96} have already analyzed a particular case 
($k=0$) of the coupling function given in Eq. (\ref{eq:w32}). The 
$k\neq0$ term included here is nevertheless very interesting because 
it allows for models with $\Phi\rightarrow\infty$ and $3+2\omega 
\rightarrow 3k/\lambda^2>0$ as $t\rightarrow0$. As shown by Serna \& 
Alimi \cite{SA96b}, a wide class of such models are able to 
significantly deviate from GR during primordial nucleosynthesis and 
still predict the observed primordial abundance of light elements. In 
addition, such models allow for baryon densities much larger than in 
GR. 

We will just analyse the models where the asymptotic functional form of the scalar 
field $\Phi$ is a power law of $\tau$: $\Phi\propto 
\tau^{\beta}$ or $\Phi\propto \tau^{- \beta}$, where $\beta>0$. However as we will see later, this $\tau$-dependance for the $\Phi$ field is found for most of the solutions.
Consequently:

(i) When $\Phi=c\tau^{-\beta}\rightarrow\infty$ at $\tau\rightarrow0$, 
we can approximate $\Phi-1\simeq\Phi =c \tau^{-\beta}$ ($c=$const.) 
and Eq. (\ref{eq:w32}) becomes

\begin{equation}
\mid\!3+2\omega\!\mid/3 \equiv |W(\tau)| = \alpha^{2}+b\tau^{\epsilon}
\label{coupl}
\end{equation}

\noindent
where $b\equiv(\lambda^2c^\delta)^{-1}$, $\epsilon\equiv\beta\delta$, 
and $\alpha^2=k/\lambda^2$ denotes the $\mid\!3+2\omega\!\mid/3$ value 
at $\tau=0$. 

(ii) When $\Phi=c\tau^{\beta}\rightarrow0$ at $\tau\rightarrow0$, we 
can approximate $1/(1-\Phi)^\delta \simeq 1+\delta\Phi = 1 + 
c\delta\tau^{\beta}$ and the coupling function is again given by Eq. 
(\ref{coupl}) but now with $b=c\delta/\lambda^2$, $\epsilon=\beta$, 
and $\alpha^2= (k+1)/ \lambda^2$.

\subsubsection{Intermediary times}

Although $\omega(\Phi)$ can be any arbitrary function of 
$\Phi$, in order to discuss the behavior of scalar-tensor cosmologies at intermediary times we must impose some conditions on the coupling function. 

Compatibility with solar-system experiments is only possible 
when the coupling function converges to the GR value 
($\omega\rightarrow\infty$) at late times. The 
simplest way of ensuring this condition is to impose that 
$\omega(\Phi)$ is a monotonic (but arbitrary) function of $\Phi$ so that it becomes infinity when $\Phi\rightarrow1$. In addition, we chose the coupling function so that it implies a scalar field ,$\Phi$, which is a 
monotonic function of $\tau$ (i.e., the variable $y$ of Eq. \ref{eq:y} does not become zero at a non-vanishing finite time). Consequently, 
the function $W(\tau)$ will be also a monotonic function of time.

An important consequence of this choice for the coupling function is 
that Eqs. (\ref{f_0}) and (\ref{W_0}) are now necessary and sufficient 
conditions to ensure the requirement $x(\tau)>0$ for any $\tau>0$. As 
a matter of fact, if $\sigma=+1$ and $y_0\leq0$, function $y(\tau)$ 
must be negative at any time and, from Eq. (\ref{eq:y'}), its absolute 
value $|y(\tau)|$ will be a monotonic decreasing function of $\tau$. 
Since Eq. (\ref{eq:f}) implies that $f(\tau)$ is always a monotonic 
increasing function, we deduce that $x(\tau)=f^2-y^2/4$ is also a 
monotonic increasing function of $\tau$. Consequently, $x(0)=0$ 
implies that $x(\tau)>0$ for any $\tau>0$.

On the other hand, if $\sigma=+1$ and $y_0>0$, the monotonic behavior 
of $W(\tau)$, together with Eqs. (\ref{eq:y'}), (\ref{eq:f}) and 
(\ref{W_0}), imply $y'(\tau)/2 = \gamma_2/(2\sqrt{|W(\tau)|} < 
\gamma_1/2 = f'(\tau)$, that is, $x'(\tau)>0$. Consequently, $x(0)=0$ 
implies that $x(\tau)>0$ for any $\tau>0$.
 
\section{Models with $3+2\omega>0$}\label{asymptotic}

The analysis of models with $3+2\omega>0$ is specially simple because, in that case, the 
variable $x$ can be written as (see Eq. \ref{f})
\begin{equation}
x=u(\tau)v(\tau)\label{factorization}
\end{equation}
with 
\begin{eqnarray}
u&\equiv& f+y/2=\frac{\gamma_1}{2}\tau+\frac{\gamma_2}{2}g(\tau)+u_0\label{u}\\
v&\equiv& f-y/2=\frac{\gamma_1}{2}\tau-\frac{\gamma_2}{2}g(\tau)+v_0\label{v}
\end{eqnarray}

Here $u_0\equiv f_0+y_0/2$, $v_0\equiv f_0-y_0/2$, while
\begin{equation}
g(\tau)\equiv\int_{0}^{\tau}\frac{d\tau}{\sqrt{\mid\!3+2\omega\!\mid/3}}
\label{g}
\end{equation}

In terms of $u$ and $v$, Eqs. (\ref{Phi(tau)})-(\ref{a(tau)}) can be expressed as
\begin{eqnarray}
\left(\frac{\Phi}{\Phi_1}\right)^{\frac{\gamma_2}{2}}&=& \left(\frac{e^{J_{u}}}{u}\right)\left(
\frac{ e^{J_{v}}}{v}\right)\label{Phi2}\\
\left(\frac{a}{a_1}\right)^{ \gamma_2}&=& \left(\frac{u}{e^{\frac{2J_u}{\gamma_1}}}\right)
\left(\frac{v}{e^{\frac{2J_v}{\gamma_1}}}\right)\label{a4}
\end{eqnarray}
with
\begin{equation}
J_{u}=\frac{\gamma_1}{2}\int_{\tau_1}^{\tau}\frac{d\overline{\tau}}{u} ;\hspace{0.5cm}
J_{v}=\frac{\gamma_1}{2}\int_{\tau_1}^{\tau}\frac{d\overline{\tau}}{v}
\label{j_uv}
\end{equation}

At early times, the coupling function can be represented by Eq. (\ref{coupl}) which, 
introduced into Eq.(\ref{g}), implies 

\begin{equation}
g(\tau)=\left\{
\begin{array}{ll}
\frac{1}{\alpha}\tau\left[ 1-\frac{b\tau^{\epsilon}}
{2\alpha^2(1+\epsilon)}+...\right] & (\alpha>0)\\
\frac{2\tau^{(2-\epsilon)/2}}{\sqrt{b}(2-\epsilon)} & (\alpha=0)
\end{array}\right. \label{gwk>0}
\end{equation}
and, hence, $u$ and $v$ are given by:

If $\alpha>0$:
\begin{eqnarray}
u &=& u_0+\frac{1}{2}[\gamma_1+\gamma_2/\alpha]\tau-
\frac{\gamma_2 b}{4\alpha^3(1+\epsilon)}\tau^{(1+\epsilon)}+...\label{uk>0}\\
v &=& v_0+\frac{1}{2}[\gamma_1-\gamma_2/\alpha]\tau+
\frac{\gamma_2 b}{4\alpha^3(1+\epsilon)}\tau^{(1+\epsilon)}-...\label{vk>0}
\end{eqnarray}

If $\alpha=0$:
\begin{eqnarray}
u &=& u_0+\frac{\gamma_1}{2}\tau+\frac{\gamma_2}{\sqrt{b}(2-\epsilon)}
\tau^{(2-\epsilon)/2}\label{u0}\\
v &=& v_0+\frac{\gamma_1}{2}\tau-\frac{\gamma_2}{\sqrt{b}(2-\epsilon)}
\tau^{(2-\epsilon)/2}\label{v0}
\end{eqnarray}
The early-time evolution of $\xi$, $a$, and $\Phi$ can be then obtained by using
Eqs. (\ref{xi}), (\ref{Phi2})-(\ref{j_uv}), and approximating the $u$ and $v$ functions by the first non-vanishing term in Eqs. (\ref{uk>0})-(\ref{v0}). Furthermore, the integration constants appearing
in this procedure must be chosen according to the criteria specified in section \ref{origin}. 

We will now analyze both the general features and early-time solutions of scalar-tensor models by considering separately the different possible values of constants $\alpha$ and $y_0$. Both aspects will be illustrated by an example solved analytically by using Eqs. (\ref{factorization})-(\ref{j_uv}) or, in some cases, numerically.

\subsection{Models with $\alpha\neq 0$ and $y_0<0$ ($\Phi'(0)<0$)}

\subsubsection{General features at arbitrary times}

Since $y_0<0$,  conditions established in Section \ref{Coupling} imply that $\Phi(\tau)$ is a monotonic decreasing function ($y(\tau)<0$) and, therefore, the scale factor $a(\tau)$ has a monotonic expansion without stationary points (see Eq. \ref{h}). In addition, the requirement $x(0)=0$ implies the existence of an initial singularity ($a(0)=0$).

These behaviors are shown in Figures 1, which present numerical solutions obtained by considering those theories defined, at any time, by
$W(\tau)=(p+q\tau)^3$ (with $\alpha^2\equiv p^3>0$, $b\equiv3p^2q>0$).

\subsubsection{Early-time solutions}

As shown in Sect. \ref{origin}, when $y_0<0$, a necessary condition 
to have the latest singularity at $\tau=0$ is that integration constants 
satisfy $f_0=-y_0/2$ (see Eq. \ref{f_0}). Then, as a first-order approximation, Eqs. (\ref{uk>0}) and (\ref{vk>0}) reduce to 

\begin{eqnarray}\label{uv}
u &=& \frac{1}{2}[\gamma_1+ \gamma_2/\alpha]\tau\\
v &=& - y_0
\end{eqnarray}

Using Eqs. (\ref{Phi2})--(\ref{j_uv}), (\ref{xi}) and (\ref{tau}), the early 
time behavior of models is then given by

\begin{mathletters}
\begin{eqnarray}
\Phi&\propto&\tau^{\frac{-2}{\gamma_1\alpha+\gamma_2}}\propto 
t^{\frac{-2}{3\alpha+1}}\label{SolIVAa}\\
a&\propto&\tau^{\frac{\alpha+1}{\gamma_1\alpha+\gamma_2}}\propto
t^{\frac{\alpha+1}{3\alpha+1}}\label{SolIVAb}\\
\xi &\propto& \tau^{-\frac{(2-
3\gamma) +\gamma_1\alpha}{2(\gamma_2+\gamma_1\alpha)}}\propto t^{-
1+\frac{3\gamma}{2}
\frac{\alpha+1}{3\alpha+1}}\label{xic}\\
t&\propto& \tau^{\frac{3\alpha+1}{\gamma_1\alpha+\gamma_2}}
\end{eqnarray}
\end{mathletters}

As expected, all models are singular ($a(0)=0$) and $\Phi(0)\rightarrow+\infty$. Thus, we can consider $\Phi\simeq\mid\!\Phi-1\!\mid$ at early-times so that, 
by identifying $\alpha=\sqrt{k}/\lambda$ and $\epsilon = 2\delta/(\gamma_1\alpha+\gamma_2)$, the coupling function (\ref{coupl}) reduces to the form of Eq. (\ref{eq:w32}).

From Eq. (\ref{xic}), we also find that the early behavior of the speed-up factor depends on the $\alpha$ parameter: $\xi$ initially becomes zero ($\xi(0)\rightarrow0$), finite ($\xi(0)\rightarrow\xi_0>0$), or infinity ($\xi(0)\rightarrow+\infty$) depending on whether $\alpha$ is smaller, equal or larger, respectively, than a critical value $\alpha_c=(3\gamma-2)/\gamma_1$.
These three kinds of models are shown in Figures 1, where we also see that the case $\xi(0)=0$ has a non-monotonic behavior.

\subsection{Models with $\alpha\neq 0$ and $y_0>0$ ($\Phi'(0)>0$)}

\subsubsection{General features at arbitrary times}

According to Eqs. (\ref{f_0}) and (\ref{W_0}), integration constants in these
models must satisfy $f_0=y_0/2$. When this condition is introduced into Eq.
(\ref{h}), we find that $a(\tau)$ presents an initial contracting phase
($h(0)<0$) provided that $W_0<1$. Since viable models imply an expanding Universe at late times (what is guaranteed by Eq. \ref{h} once $W$ increases and becomes greater than unity), models with $W_0<1$ necessarily have a minimum in the
scale factor and, hence, they are non-singular. 

In the opposite, when $W_0\geq1$, the Universe evolution starts with 
an expanding phase ($h>0$) which will continue at any later time. As a 
matter of fact, the requirement $x(\tau)>0$ together with Eq. 
(\ref{f}) imply that $f(\tau)>y(\tau)/2$ for any $\tau>0$. 
Consequently, $h(\tau)$ is always positive.

\subsubsection{Early-time solutions}

Using $f_0=y_0/2$ the first-order approximation of Eqs. (\ref{uk>0}) and (\ref{vk>0}) reduce to

\begin{eqnarray}
u &=& y_0\\
v&=&\left\{
\begin{array}{ll}
\frac{1}{2}[\gamma_1 - \gamma_2/\alpha]\tau & (\alpha>\gamma_2/\gamma_1)\\
\frac{\gamma_{1}^{3}b}{4\gamma_{2}^{2}(1+\epsilon)}\tau^{(1+\epsilon)}
& (\alpha=\gamma_2/\gamma_1)
\end{array}\right.
\end{eqnarray}
where
\begin{equation}
\alpha\ge\gamma_2/\gamma_1\label{condition2}
\end{equation}
is required according to Eq. (\ref{W_0}).

Consequently, when $\alpha>\gamma_2/\gamma_1$, the early-time behavior of the dynamical fields and variables is 

\begin{mathletters}
\begin{eqnarray}
\Phi&\propto&\tau^{\frac{2}{\gamma_1\alpha-\gamma_2}}\propto 
t^{\frac{2}{3\alpha-1}}\label{SolIVBa}\\
a&\propto&\tau^{\frac{\alpha-1}{\gamma_1\alpha-\gamma_2}}\propto
t^{\frac{1-\alpha}{1-3\alpha}}\label{SolIVBb}\\
\xi &\propto& \left\{
\begin{array}{ll}
\tau^{-\frac{\gamma_1\alpha-(2-3\gamma)}{2[\gamma_1\alpha-
\gamma_2]}}\propto t^{-1+\frac{3\gamma}{2}\frac{1-
\alpha}{1-3\alpha}} & (\alpha\neq 1)\\
\mbox{const.}>0 & (\alpha=1)
\end{array}\right.\\
t&\propto& \tau^{-\frac{1-3\alpha}{\gamma_1\alpha-\gamma_2}}
\end{eqnarray}
\end{mathletters}
\noindent while, when $\alpha=\gamma_2/\gamma_1$, we find that 
solutions have exponential behaviors which are not compatible with 
a coupling function $3+2\omega(\Phi)$ as that assumed in 
Eq. (\ref{eq:w32}).
 

Since $\alpha$ must satisfy Eq. (\ref{condition2}), the above 
solutions imply that $\Phi(0)=0$. Consequently, by identifying $\alpha^2 = 
(k+1)/\lambda^2$ and $\epsilon = 2/(\gamma_1\alpha-\gamma_2)$, the coupling function (\ref{coupl}) can be expressed in the form given by Eq. (\ref{eq:w32}). 

The scale and speed-up factors have instead a more complicated 
behavior depending on the $\alpha$ value:

{\it i)} If $\alpha>1$ ($\omega(0) > 0$), the scale factor vanishes at 
$\tau=0$ while $\xi$ diverges to $+\infty$. Models are then singular. 

{\it ii)} If $\alpha<1$ ($\omega(0) < 0$), then $a(0)$ diverges to 
$+\infty$ and $\xi$ diverges to $-\infty$. Models are then non-singular. 

{\it iii)} $\alpha=1$ ($\omega(0) = 0$), the scale and speed-up 
factors have a finite non-vanishing value at $\tau=0$ while 
$\Phi(0)=0$. In this case, the $a$ and $\Phi$ derivatives are both 
positive at $\tau=0$ and, hence, a negative $\tau$ value exists where 
the scale factor has a singular point and $\Phi<0$. We then note that 
our choice of the origin of times ensures that gravitation is always 
attractive for $\tau>0$. However, in this case, an alternative choice 
of the origin of times could also be the requirement $a(0)=0$ 
(implying a repulsive gravitation at early times). 

All these behaviors are shown in Figures 2, which represent the {\it 
analytical} solutions obtained when the coupling function is 
given by $(3+2\omega)/3=\alpha^2+b\tau$ ($\alpha=1$) (see appendix B), as well as the numerical solutions obtained for $(3+2\omega)/3=\alpha^2+b\tau^{1/2}$ ($\alpha\neq1$) . 
We also see from these figures that models with $\alpha<1$ can present 
a non-monotonic behavior of the speed-up factor.

\subsection{Models with $\alpha\neq 0$ and $y_0=0$ ($\Phi'(0)\geq0$)}

\subsubsection{General features at arbitrary times}

According to Eq. (25), the expanding or contracting behavior of the scale factor, $a(\tau)$, is determined by the sign of $[f - y/(2\sqrt{W})]$. 

Taking into account that $W(\tau)$ is a monotonic increasing function ($1/\sqrt{W} < 1/\sqrt{W_0}$) and that integration constants must satisfy $f_0=y_0/2=0$, Eq. (\ref{eq:y}) implies that $y(\tau)<\gamma_2\tau/\sqrt{W_0}$. Consequently, $y/\sqrt{W}< \gamma_2\tau/W_0$ and, using Eq. (\ref{eq:f}) (with $f_0=0$), we obtain
\begin{eqnarray}
 [f - y/(2\sqrt{W})]>(\gamma_1-\gamma_2/W_0)\tau/2\label{ineq}
\end{eqnarray}

\noindent
which implies two different situations in the time evolution of $a(\tau)$, depending on the initial value of $W(\tau)$:

{\it i)} $W_0 \geq \gamma_2/\gamma_1$: in this case, $[f - y/(2\sqrt{W})]> 0$ for any $\tau>0$, and there exists a undefined expansion. 

{\it ii)} $W_0 < \gamma_2/\gamma_1$: A similar argument as that used in the paragraph preceding Eq. (\ref{ineq}) implies that for any time previous to that, $\tau_1$, in which $W$ reaches the value $W_1 \equiv \gamma_2/\gamma_1$:

\begin{equation} 
[f - y/(2\sqrt{W})]<0
\end{equation}

Consequently, there exists an initial phase of contraction in the scale factor (for $\tau<\tau_1$). Since $f(\tau) > y(\tau)/2$ ($x(\tau) > 0$), Eq. (25) implies an expansion process when $W$ becomes larger than unity. Then, the scale factor must present a minimum (Fig. 3b), in the time interval between $W = \gamma_2/\gamma_1$ and $W = 1$.

\subsubsection{Early-time solutions}

According to Eqs. (\ref{f_0}) and (\ref{W_0}), integration constants must satisfy $f_0=y_0/2=0$ and $\alpha\ge\gamma_2/\gamma_1$. Consequently, the $u$ and $v$ functions can be approximated by

\begin{eqnarray}
u &=& \frac{1}{2}(\gamma_1+\gamma_2/\alpha)\tau\\
v &=& \left\{
\begin{array}{ll}
\frac{1}{2}(\gamma_1-\gamma_2/\alpha)\tau & (\alpha>\gamma_2/\gamma_1)\\
\frac{\gamma_{1}^{3}b}{4\gamma_{2}^{2}(1+\epsilon)}\tau^{(1+\epsilon)} &
(\alpha=\gamma_2/\gamma_1)
\end{array}\right.
\end{eqnarray}

When $\alpha>\gamma_2/\gamma_1$, the early-time behavior of models is then given by:

\begin{mathletters}
\begin{eqnarray}
\Phi&\propto& \tau^{4\gamma_2/A}
\propto t^{4\gamma_2/B}\\
a&\propto& \tau^{2(\gamma_1\alpha^2-\gamma_2)/A} \propto 
t^{2(\gamma_1\alpha^2-\gamma_2)/B}\label{SolIVCb}\\
\xi &\propto& \left\{
\begin{array}{ll}
\tau^{-2\gamma_2/A} 
\propto t^{-2\gamma_2/B} & (\alpha^2\ne\gamma_2/\gamma_1)\\
\tau\propto t
& (\alpha^2=\gamma_2/\gamma_1)
\end{array}\right.\\
t &\propto& \tau^{B/A}\\
A &=& \gamma_{1}^{2}\alpha^2-\gamma_{2}^{2};\hspace{0.5cm}
B=2\gamma_2+3\gamma(\gamma_1\alpha^2-\gamma_2)
\end{eqnarray}
\end{mathletters}

\noindent
while, when $\alpha=\gamma_2/\gamma_1$, solutions are 
exponentials which are not compatible with Eq. (\ref{eq:w32})

We then find that all models imply $\Phi(0)\rightarrow0$. Consequently, by identifying $\alpha^2 = (k+1)/\lambda^2$ and $\epsilon = 4\gamma_2/(\gamma_1^2\alpha^2-\gamma_2^2)$, the coupling function (\ref{coupl}) can be expressed in the form given by Eq. (\ref{eq:w32}). The scale and speed-up factors have instead a wider variety of early-time behaviors. If $\alpha^2>\gamma_2/\gamma_1$ ($\epsilon<2$), models are singular ($a(0)\rightarrow0$ and $\xi(0)\rightarrow+\infty$). If $\alpha^2<\gamma_2/\gamma_1$, models are non-singular with $a(0)\rightarrow+\infty$ and $\xi(0)\rightarrow-\infty$. Finally, if $\alpha^2=\gamma_2/\gamma_1$ ($\epsilon=2$), models are again non-singular but now with $a(0)\rightarrow$const. (a minimum) and $\xi(0)\rightarrow0$. These behaviors are shown in Figures 3, which represent the analytical exact solutions for $W=\alpha^2+b\tau$ (case $\alpha^2>\gamma_2/\gamma_1$ of appendix B), and the numerical solutions for $W=\alpha^2+b\tau^2$ ($\alpha^2=\gamma_2/\gamma_1$) and $W=(\alpha+q\tau^3)^2$ ($\alpha^2<\gamma_2/\gamma_1$).

\subsection{Models with $\alpha=0$}

\subsubsection{General features at arbitrary times}

According to Eq. (34) these models are only compatible with the initial condition $f_0=-y_0>0$. Consequently, $\Phi(\tau)$ is a monotonic decreasing function and Eq. (\ref{h}) implies that the scale factor $a(\tau)$ increases indefinitely from a vanishing value ($x(0)=0$) at the origin of times. This behavior is shown in Figures 4, which display the numerical solutions obtained when $(3+2\omega)/3=b\tau(1+c\tau^{1/2})^3$ ($b,c>0$) at any time.

\subsubsection{Early-time solutions} 

When $3+2\omega$ vanishes as $\tau\rightarrow0$, the coupling function can be expressed as
\begin{equation}
\mid\!3+2\omega\!\mid/3 =b\tau^{\epsilon}\label{w0}
\end{equation}
with $b>0$ and $0<\epsilon<2$. In this case, the first-order approximation of $u$ and $v$ functions will be given by (see Eqs. \ref{u0} and \ref{v0}):
\begin{mathletters}
\begin{eqnarray}
u&=&\frac{\gamma_2}{\sqrt{b}(2-\epsilon)}
\tau^{(2-\epsilon)/2}\\
v&=& v_0=-y_0
\end{eqnarray}
\end{mathletters}

Then, Eqs. (\ref{Phi2})--(\ref{j_uv}) give

\begin{mathletters}\label{solutw0}
\begin{eqnarray}
\Phi &\propto& \tau^{-2/(\gamma_2+\delta)}\propto 
t^{-2/(1+\delta)}\label{Phiw0}\\
a &\propto& \tau^{1/(\gamma_2+\delta)}\propto 
t^{1/(1+\delta)}\\
\xi &\propto&\tau^{-\frac{\gamma_2+2(\delta-1)}{2(\gamma_2+\delta)}}\propto 
 t^{-\frac{\gamma_2+2(\delta-1)}{2(1+\delta)}}\\
t &\propto& \tau^{\frac{1+\delta}{\gamma_2+\delta}} 
\end{eqnarray}
\end{mathletters}
where
\begin{equation}
\delta \equiv \frac{\gamma_2\epsilon}{2-\epsilon}
\end{equation}
corresponds to the exponent of $|\Phi-1|$ appearing in Eq. (\ref{eq:w32}).

As expected, we find that all models are singular and $\Phi(0)\rightarrow+\infty$. The early behaviour of $\xi$ depends instead of the $\delta$ value. If $\delta>
\delta_c\equiv (3\gamma-2)/2$, we find that $\xi(0)\rightarrow+\infty$. If $\delta=\delta_c$, $\xi(0)$ has a finite positive value. Finally, if $\delta<\delta_c$, $\xi$ vanishes at $\tau=0$ and presents a non-monotonic behavior (see Figures 4). In any case, since $\Phi$ diverges to infinity at $\tau\rightarrow0$, we can approximate $\Phi\simeq\Phi-1$ and Eq. (\ref{w0}) becomes the coupling function given by Eq. (\ref{eq:w32}) with $k=0$.

\section{Models with $3+2\omega<0$}\label{asymptotic1}

According to Eq. (\ref{f}), models with $W < 0$ are characterized by the fact that $x(\tau) = f^2 + y^2/4$. Consequently, $x$ has always a positive value which can be initially vanishing only if integration constants are chosen so that they
satisfy the condition given by Eq. (\ref{f0s<0}): $f_0=y_0=0$

 Since the factorization (\ref{factorization}) is not further possible, solutions to the scalar-tensor cosmological equations can only be obtained by applying the general procedure described in Eqs. (\ref{Phi(tau)})-(\ref{xi}). The early-time expressions can be deduced by considering Eq. (\ref{coupl}) and approximating $y(\tau)$ by the first non-vanishing term of:
 
\begin{equation}\label{yW<0}
y(\tau)=y_0 - \gamma_2\left\{
\begin{array}{ll}
(\tau/\alpha)[1-\frac{b\tau^\epsilon}{2\alpha^2(1+\epsilon)}+...], &
(\alpha>0)\\
\frac{2\tau^{(2-\epsilon)/2}}{\sqrt{b}(2-\epsilon)}, & (\alpha=0)
\end{array}\right.
\end{equation}

\noindent
where integration constants must be chosen according to the criteria established in Section \ref{origin}

\subsection{Models with $\alpha \neq 0$ and $y_0 < 0$, ($\Phi'(0) < 0$ )}
\label{VA}

\subsubsection{General features at arbitrary times}

Although we already know that all these models have a stationary point at $\tau=0$ (see Eq. \ref{minimum}), we will now analyze the conditions needed so that such a stationary point corresponds to a minimum in $a(\tau)$. 

From the stationary point condition (Eq. \ref{minimum}), $f_0 = y_0/(2\sqrt{|W_0|})]<0$, the function $f(\tau)$ is initially negative, but it becomes positive for any $\tau>\tau_1\equiv2|f_0|/\gamma_1$. Consequently, we can ensure that the scale factor expands for $\tau>\tau_1$ (see Eq. \ref{h}) and, therefore, it presents a minimum value at $\tau=0$ provided that:
\begin{equation}
{f'}_0=\gamma_1/2>{[y/(2\sqrt{|W|})]'}_0\label{mincond}
\end{equation}

If the early-time behavior of $W(\tau)$ is represented by Eq. (\ref{coupl}), the above condition is equivalent to $\epsilon\geq1$, with the additional constraint $b|y_0|/2 < \alpha(\gamma_1\alpha^2 + \gamma_2)$ for $\epsilon = 1$.

\subsubsection{Early-time solutions}

Since $x(0) = f_{0}^{2} + y_{0}^{2}/4 > 0$, condition (\ref{minimum}) implies that $a(0)$ and $\Phi(0)$ are finite and non-vanishing, while $\xi(0)=0$. Consequently, as a first-order approximation:

\begin{equation}
\Phi\rightarrow \mbox{const.}>1,\hspace{0.2cm}
a\rightarrow \mbox{const.}>0,\hspace{0.2cm}
\xi\rightarrow 0, ,\hspace{0.2cm}
t\propto\tau
\end{equation}

Using Eq. (\ref{yW<0}) to obtain a higher approximation, we find: 

\begin{mathletters}\label{SolVA}
\begin{eqnarray}
\Phi &\simeq& \Phi(0)(1+c_1\tau)\label{70a}
\\
a &\simeq& a(0) (1+c_2\tau^2)\\
\xi &\propto& \tau\\
t &\propto& \tau
\end{eqnarray}
\end{mathletters}
where $a(0)>0$, $\Phi(0)>1$ and 
\begin{mathletters}
\begin{eqnarray}
c_1&\equiv&\frac{4\alpha}{y_0(1+\alpha)^2}\\
c_2&\equiv&
\frac{\gamma_1\alpha^2+\gamma_2+y_0b/2\alpha}{y_{0}^{2}(1+\alpha^2)}, \;
(\epsilon=1)
\end{eqnarray}
\end{mathletters}

Models are then non-singular and the scalar field takes a finite non-vanishing value at early-times (see Figure 5). It must be noted that, from Eq. (\ref{70a}), the only $\epsilon$ value consistent with Eq. (\ref{eq:w32}) is $\epsilon=1$.

\subsection{Models with $\alpha \neq 0$ and $y_0 > 0$, ($\Phi'(0)>0$)}

\subsubsection{General features at arbitrary times}

According to the conditions established in Sect. \ref{origin}, $\Phi(\tau)$ is a monotonic increasing function with an initial value $0 < \Phi(0) < 1$ .

 On the other hand, since $f(\tau)$ and $|W(\tau)|$ are monotonic increasing functions while $y(\tau)$ decreases monotonically, Eq. (\ref{minimum}) implies

\begin{equation}
[ f - y / (2\sqrt{|W|})] > [f_0 - y_0/ (2\sqrt{|W_0|}) ] = 0
\end{equation}

\noindent
and, according to Eq. (\ref{h}), $a(\tau)$ then expands at any time from an initial minimum value $a(0) > 0$ (see Fig. 6)

\subsubsection{Early-time solutions}

By following the same procedure as in Sect. \ref{VA}, we obtain again the solutions (\ref{SolVA}), but now with $0 < \Phi(0)<1$.

\subsection{Models with $\alpha \neq 0$ and $y_0 = 0$, ($\Phi'(0) < 0$ )}

\subsubsection{General features at arbitrary times}

From the initial condition (\ref{f0s<0}) , $f_0=y_0= 0$, we find that $f(\tau) > 0$ and $y(\tau)<0$ for any $\tau>0$ (strictly). Consequently, $\Phi(\tau)$ decreases monotonically while, from Eq. (\ref{h}), $a(\tau)$ expands indefinitely. This behavior is shown in Figure 7, which displays the numerical solutions obtained in models defined by $|W(\tau)|=\alpha^2+b\tau$ at any time.

\subsubsection{Early-time solutions}

Using $f_0=y_0=0$, the $y(\tau)$ function can be approximated (see Eq. \ref{yW<0}) by

\begin{equation}
y(\tau)=-(\gamma_2/\alpha)\tau
\end{equation}

Thus, we obtain

\begin{mathletters}
\begin{eqnarray}
\Phi &\propto& \tau^{-4\gamma_2/A}\propto t^{-4\gamma_2/B}\\
a &\propto& \tau^{2(\gamma_1\alpha^2+\gamma_2)/A}\propto
t^{2(\gamma_1\alpha^2+\gamma_2)/B}\\
\xi &\propto& \tau^{2\gamma_2/A}\propto t^{2\gamma_2/B}\\
t &\propto& \tau^{B/A}\\
A &\equiv& \gamma_{1}^{2}\alpha^2+\gamma_{2}^{2}, \hspace{0.3cm}
B \equiv 3\gamma\gamma_1(1+\alpha^2)-8
\end{eqnarray}
\end{mathletters}
 
Models are then singular (with $a=a'=0$) and $\Phi(0)\rightarrow+\infty$. 

It must be noted that, for $\gamma$ values implying $B<0$, the above 
early-time solutions correspond to the limit $t\rightarrow-\infty$ 
instead of $t\rightarrow0$. This is also true for any other asymptotic 
solution where $t$ is proportional to a negative power of $\tau$.

\subsection{Models with $\alpha=0$}

\subsubsection{General features at arbitrary times}

As we have seen in Sect. \ref{origin}, when $3+2\omega(0)$ vanishes 
from negative values, the only possible choice of integration 
constants is $f_0=y_0=0$ ($\Phi'(0)\leq0$). Just like in the previous 
subsection, this choice ensures that $f(\tau) > 0$ and $y(\tau) < 0$ 
for any $\tau>0$. Consequently, $\Phi(\tau)$ is a monotonic decreasing  
function while, from Eq. (\ref{h}), $a(\tau)$ expands indefinitely 
from an initial singularity ($x(0)=0$).  This behavior is shown in 
Figures 8 through the analytical exact solutions obtained when 
$|W(\tau)|=b\tau^\epsilon$ at any time (see also Appendix B).

\subsubsection{Early-time solutions} 

Using Eq. (\ref{yW<0}) with $f_0=y_0=0$, the $y(\tau)$ function
can be approximated by:

\begin{equation}
y(\tau)=-\frac{2\gamma_2}{(2-\epsilon)\sqrt{b}}\tau^{(2-\epsilon)/2}
\hspace{0.3cm} (\epsilon<2)
\end{equation}

Thus, we obtain: 

\begin{mathletters}
\begin{eqnarray}
\Phi &\propto& \tau^{-4/A}\propto t^{-4/B}\\
a &\propto& \tau^{2/A}\propto t^{2/B}\\
\xi &\propto& \tau^{2(1-\delta)/A}\propto t^{2(1-\delta)/B}\\
t &\propto& \tau^{B/A}\\
A &\equiv& \gamma_2+2\delta,\hspace{0.3cm} B\equiv 2\delta-\gamma_2+2
\end{eqnarray}
\end{mathletters}

\noindent
where $\delta\equiv\gamma_2\epsilon/[2(2-\epsilon)]$ is the exponent appearing in Eq. (\ref{eq:w32}).

Models are then singular and the scalar field decreases from $\Phi(0)
\rightarrow+\infty$. The initial value of the speed-up factor depends instead on the $\delta$ value. If $\delta > 1$ , $\xi(0)$ diverges to $+\infty$, (Fig. 8a); if $\delta = 1$, $\xi(0)$ has a finite non-vanishing value; finally, if $\delta < 1$ , $\xi(0)$ becomes zero, (Fig 8b).

Table 1 summarizes all the early-time asymptotic solutions that we have found in sections \ref{asymptotic}-\ref{asymptotic1} for $\gamma=0$ 
(false-vacuum models) and $\gamma=1$ (matter-dominated models). This table also contains the early-time 
solutions for vacuum and radiation-dominated models ($\gamma=4/3$) obtained in our previous work 
\cite{SA96a}. The way in which these results can be applied to derive the asymptotic solutions of other scalar-
tensor theories (not strictly defined by Eq. \ref{eq:w32}) is illustrated in the Appendix A.

\section{Late Time solutions}\label{asymptotic2}

We will analyze in this section the way in which scalar-tensor 
theories converge towards GR at late times. We note however that, 
although solar-system experiments require in fact that any viable 
scalar-tensor theory must converge towards GR during the 
matter-dominated era, such a condition is not compulsory for earlier 
epochs in the universe evolution. Serna \& Alimi \cite{SA96b} have in 
fact shown that some scalar-tensor theories are able to predict the 
right primordial abundances of light elements even when such theories 
are very different from GR during primordial nucleosynthesis. 
Consequently, at the end of the inflationary epoch, a scalar-tensor 
theory can deviate considerably from GR and still be compatible with 
all astronomical observations. 

When convergence to GR ($3+2\omega\rightarrow\pm\infty$) is imposed at 
late-times ($\tau\rightarrow\infty$), Eqs. (\ref{eq:y'}), (\ref{f}) 
and (\ref{eq:f}) imply $y'\rightarrow0$ and $x(\tau)\simeq 
\gamma_{1}^{2}\tau^2/4$. The integration method described in Sect. 
\ref{sec:method} then lead to solutions which, as a first-order 
approximation, are similar to those found in GR:

\begin{mathletters}
\begin{eqnarray}
\Phi & \simeq & \xi \simeq 1\\
a & \propto & \tau^{2/\gamma_1}\\
t & \propto & \left\{
\begin{array}{ll}
\ln(\tau) & (\gamma=0)\\
\tau^{3\gamma/\gamma_1} & (\gamma\neq0)
\end{array}\right.
\end{eqnarray}
\end{mathletters}

A higher-order approximation to the $\Phi$ and $\xi$ solutions can be found by considering (see Eq. \ref{coupl}) that, in the limit $\Phi\rightarrow1$, the asymptotic form of the coupling function is 
 
\begin{equation}
|3+2\omega(\Phi)|/3=\lambda^{-2}/|\Phi-1|^{\delta}=b\tau^{\epsilon}\label{w32:conv}
\end{equation}
with $\lambda>0$, $\delta>0$, $b>0$, and $\epsilon>0$.

Using Eq. (\ref{w32:conv}), the $y(\tau)$ function is given by
\begin{equation}
y(\tau)=y(\tau_1)+\frac{2\sigma\gamma_2}{b(2-\epsilon)}
(\tau-\tau_1)^{(2-\epsilon)/2}; \hspace{0.3cm}(\epsilon\neq2)\label{yconv}
\end{equation}
where $\tau_1$ denotes the time at which Eq. (\ref{w32:conv}) starts to be a good approximation for the coupling function. The $\epsilon=2$ value in Eq. (\ref{yconv}) has been excluded because it leads to models which are incompatible with Eq. (\ref{eq:w32}).

In the limit $\tau\rightarrow\infty$, we can find two possibilities for the $y$ function: {\it a)} $y(\tau)$ converges towards a finite value, $y_c$, and {\it b)} $y(\tau)$ diverges to $\pm\infty$. We will now analyze separately the solutions corresponding to these two cases.

\subsection{Case $y\rightarrow y_c$, ($1<\delta<2$)}

 Since $\Phi\simeq1$, $y(\tau)\simeq y_c$ and $x(\tau)\simeq \gamma_{1}^{2}\tau^2/4$, Eqs. (\ref{y}), (\ref{x}) and (\ref{w32:conv}) imply:
\begin{equation}
\frac{\Phi'}{\Phi} = \frac{y}{x\sqrt{|W|}}\simeq 
\frac{4y_c}{\gamma_{1}^{2}\sqrt{b}}\tau^{-(2+\epsilon/2)}
\end{equation}
which, together with Eq. (\ref{xi}), leads to:
\begin{mathletters}
\begin{eqnarray}
\Phi &=& 1 - \mbox{sign}(y_c)c\tau^{-2/(2-\delta)}\\
\xi &=& 1+\sigma[y_{c}^{2}/(2\gamma_{1}^{2})]\tau^{-2}
\end{eqnarray}
\end{mathletters}
where $1<\delta<2$ ($\epsilon>2$) and
\begin{mathletters}
\begin{eqnarray}
\delta &=& 2\epsilon/(2+\epsilon)\\
\lambda^{-2} &=& bc^\delta\\
c &\equiv& [2\lambda(2-\delta)|y_c|/\gamma_{1}^{2}]^{2/(2-\delta)}
\end{eqnarray}
\end{mathletters}

We then find that $\Phi$ converges towards the GR value from below ($\Phi\rightarrow1^{-}$) or above ($\Phi\rightarrow1^{+}$) depending on the sign (positive or negative, respectively) of $y_c$. On the other hand, $\xi$ converges to unity from above ($\xi\rightarrow1^{+}$) or below ($\xi\rightarrow1^{-}$) depending on the sign (positive or negative, respectively) of $3+2\omega$.

\subsection{Case $y\rightarrow \pm\infty$ ($\delta=1$)}

In this case, $\Phi\simeq1$, $y(\tau)\propto 2\sigma\gamma_2/[b(2-\epsilon)] \tau^{(2-\epsilon)/2}$ 
(see Eq. \ref{yconv}) and $x(\tau)\simeq \gamma_{1}^{2}\tau^2/4$. 
Consequently, Eqs. (\ref{y}), (\ref{x}) and (\ref{w32:conv}) imply:

\begin{mathletters}\label{solconv1a}
\begin{eqnarray}
\Phi &=& 1 - \sigma c\tau^{-\epsilon}\\
\xi &=& 1-\sigma(c/4) \tau^{-\epsilon} 
[(\epsilon-1)/(2-\epsilon)](4-\gamma_1\epsilon)
\tau^{-\epsilon}
\end{eqnarray}
\end{mathletters}
with $\delta=1$, $1/\lambda^2=bc$, and
\begin{equation}
\epsilon=1\pm[1-(8\lambda^2\gamma_2/\gamma_{1}^{2})]^{1/2}; \;
(8\lambda^2\leq \gamma_{1}^{2}/\gamma_2)\label{solconv1b}
\end{equation}

We then find that $\Phi$ converges towards the GR value from below ($\Phi\rightarrow1^{-}$) or above ($\Phi\rightarrow1^{+}$) depending on the sign (positive or negative, respectively) of $3+2\omega$. On the other hand, $\xi$ converges to unity from below ($\xi\rightarrow1^{-}$) or above ($\xi\rightarrow1^{+}$) depending on the sign (positive or negative, respectively) of $\sigma(\epsilon-1)(4-\gamma_1\epsilon)$.

\section{Conclusions} \label{sec:conclus}

We have presented in this paper a method to derive exact solutions for $K=0$ scalar-tensor cosmologies with an arbitrary 
$\omega(\Phi)$ function and satisfying the general perfect fluid state 
equation $P=(\gamma-1)\rho c^2$,  where $\gamma$ is a 
constant and $0\leq\gamma\leq2$. This procedure has some common aspects with the indirect method 
previously proposed by Barrow and Mimoso \cite{Barrow94}. In particular, solutions are not directly found 
through the coupling function $\omega(\Phi)$ which defines each specific scalar-tensor theory, but by means of 
a 'generating function' and a suitable change of variables. A given choice of the generating function produces, 
after completely solving the field equations, a particular form of $\omega(\Phi)$. Therefore, the specific class of 
scalar-tensor theories under study is not known 'a priori', but only 'a posteriori'. 

Unlike the previous indirect method, the non-linear transformation of variables used in our 'semi-indirect' 
procedure allows us to take the time dependence of the coupling function itself as the generating function. This 
has the obvious advantage that much of the main properties characterizing the functional form of 
$\omega(\Phi)$ can be easily known 'a priori', what results specially useful to analyze the asymptotic behavior of 
scalar-tensor cosmologies.

Using this procedure, we have supplied a comprehensive study of asymptotic cosmological solutions in the 
framework of a wide class of homogeneous and isotropic theories. We have also described the qualitative behavior of models
at intermediary times and, for some particular
scalar-tensor theories, we have obtained solutions which are exact at any time characterized by a constant $\gamma$ value. 
Our analysis then covers all the main epochs 
in the cosmic history (inflationary, radiation- and matter-dominated models) and, therefore, it extends other works 
(specially those of Refs. \cite{SA96a} and \cite{BP96}) also devoted to a systematic study of scalar-tensor 
models. All the early-time solutions found in this paper are summarized in Table 1. We find that singular models with $3+2\omega>0$ and $y_0\ne0$ have always early-time solutions for $a$ and $\Phi$ which are independent of $\gamma$. Consequently, in these models, all the epochs in the universe evolution have the same early behavior. 

We finally note that the solutions obtained in the present paper, together with those of other previous works 
\cite{SA96a,BP96}, enable complete cosmological histories to be constructed through initial vacuum (or false 
vacuum), radiation, matter, and final vacuum-dominated eras. The wide diversity of possible scalar-tensor 
cosmological models can be restricted from additional constraints like those obtained from the primordial 
abundance of light elements \cite{SA96b}, but also from their implications on the initial spectrum of density 
perturbations \cite{GBW}, and other strong-field tests (e.g., Ref. \cite{GB}).

\acknowledgments

This work was partially supported by the Comisi\'on Interministerial 
de Ciencia y Tecnolog\'{\i}a (Project No. ESP96-1905-E), Spain.

\section*{appendix A: Connection with other scalar-tensor theories}

The asymptotic solutions found in this paper are not restricted to 
scalar-tensor theories with a coupling function $3+2\omega$ strictly 
given by Eq. (\ref{eq:w32}). Such asymptotic solutions hold for any 
arbitrary form of $3+2\omega$ admitting (at early or late times) a 
series approximation as that expressed by Eq. (\ref{eq:w32}).

Let us consider, for example, a scalar-tensor theory defined at any 
time by 

\begin{equation}
3+2\omega=\frac{-B}{\ln\left[ (\Phi+1)/2\right]}\label{w_example}
\end{equation}
with $B>0$ and $\Phi\in(0,1)$. At early times ($\Phi\rightarrow0$), we can write $1/(1-\Phi)\simeq 1+\Phi$ 
and, hence, the Taylor approximation of Eq. (\ref{w_example}) is

\begin{equation}
3+2\omega=\frac{B}{\ln^2(2)}[\ln(2)-1 + \mid\! 1-\Phi\!\mid^{-1}]
\end{equation}
which has the form given by Eq. (\ref{eq:w32}) with $\lambda^2=3\ln^2(2)/B$, $k=\ln(2)-1$ and $\delta=1$. 
Consequently, $1/\alpha=\lambda/\sqrt{k+1}=\sqrt{3\ln(2)/B}$, and we see from Table 1 or Eqs. (\ref{SolIVBa})-(\ref{SolIVBb}) that its early solutions at any 
epoch are 

\begin{equation}
a\propto t^{\frac{\alpha-1}{3\alpha-1}} ;\hspace{0.5cm}
\Phi\propto t^{\frac{2}{3\alpha-1}}
\end{equation}
Models are then singular provided that $\alpha>1$ ($B>3\ln(2)$) while, otherwise, they are non-singular.

At late times ($\Phi\rightarrow1$, $\gamma=1$), we have $\ln\left[ (\Phi+1)/2\right]\simeq -(1-\Phi)/2$ and
\begin{equation}
3+2\omega=2B \mid\! 1-\Phi\!\mid^{-1}
\end{equation}
which is again of the form given by Eq. (\ref{eq:w32}) with $\lambda^2=3/(2B)$, $k=0$ and $\delta=1$. According to Eqs. 
(\ref{solconv1a})-(\ref{solconv1b}), this theory converges at 
$t\rightarrow\infty$ towards the GR solutions provided that 
$\lambda^2\leq 9/8$, that is, provided that $B\geq4/3$.

In the same way, much other theories (as the late-time behavior of all the classes of theories defined in 
\cite{BP96}) are asymptotically described by the solutions obtained in this paper.
 
\onecolumn
\section*{appendix B: Analytical exact solutions}
The exact solutions obtained by applying the method described in this paper to those theories defined by $(3+2\omega)/3=\alpha^2+b\tau^{\epsilon}$ ($\alpha\ge0$, $b>0$) are:

1) If $f_0=y_0/2\geq0$, $\epsilon=1$, $v_0=0$, and $\sigma=+1$

\begin{mathletters}
\begin{eqnarray}
\Phi &\propto& \left[\frac{z-\alpha}{z+\alpha - 2\gamma_2/\gamma_1}
\right]^{\frac{2}{\gamma_1\alpha-\gamma_2}}
\left\{\begin{array}{ll}
\left(\frac{z-z_1}{z-z_2}\right)^{-2/\sqrt{R}} & R>0\\
e^{-\frac{4}{\sqrt{|R|}}\mbox{atan}\left(\frac{\gamma_1 z+\gamma_2}
{\sqrt{|R|}}\right)} & R<0\\
e^{\frac{4}{\gamma_1 z+\gamma_2}} & R=0
\end{array}\right.\\
a &\propto& (z-\alpha)^{\frac{\alpha-1}{\gamma_1\alpha-\gamma_2}}
[z+\alpha-2\gamma_2/\gamma_1]^{\frac{\alpha-(2-3\gamma)/\gamma_1}
{\gamma_1\alpha-\gamma_2}}
\left\{\begin{array}{ll}
(z-z_1)^{\frac{1+2/\sqrt{R}}{\gamma_1}}(z-z_2)^{\frac{1+2/\sqrt{R}}{\gamma_1}} & R>0\\
u^{1/\gamma_1}
e^{\frac{4}{\gamma_1\sqrt{|R|}}}\mbox{atan}\left(\frac{\gamma_1 z+\gamma_2}
{\sqrt{|R|}}\right) & R<0\\
u^{1/\gamma_1}
e^{-\frac{4}{-\gamma_1(\gamma_1 z+\gamma_2)}} & R=0
\end{array}\right.
\end{eqnarray}
\end{mathletters}
where
\begin{mathletters}
\begin{eqnarray}
u &\equiv& (\gamma_1/2)\tau+(\gamma_2/b)[\sqrt{\alpha^2+b\tau}-\alpha]+u_0\\
z &\equiv& \sqrt{\alpha^2+b\tau}\\
z_1 &\equiv& \frac{-\gamma_2+\sqrt{R}}{\gamma_1}\\
z_2 &\equiv& \frac{-\gamma_2-\sqrt{R}}{\gamma_1}\\
R &\equiv& [\gamma_1\alpha+\gamma_2]^2 - 2b\gamma_1 u_0
\end{eqnarray}
\end{mathletters}

2) If $y_0=0$, $\epsilon<2$, $\alpha^2=0$ and $\sigma=-1$
\begin{mathletters}
\begin{eqnarray}
\Phi &=& [1+C\tau^{-\epsilon}]^{\frac{2(2-\epsilon)}{\gamma_2\epsilon}}\\
a &=& (\gamma_1\tau/2)^{2/\gamma_1}[1+C\tau^{-\epsilon}]^{
\frac{\gamma_1\epsilon-4}{\gamma_1\gamma_2\epsilon}}\\
\xi &=& [1+C\tau^{-\epsilon}]^{\frac{\gamma_1\epsilon-4}{2\gamma_2\epsilon}}
- C\frac{\gamma_1\epsilon-4}{2\gamma_2}\tau^{-\epsilon}
[1+C\tau^{-\epsilon}]^{\frac{\gamma_1\epsilon-4}{2\gamma_2\epsilon}-1}
\end{eqnarray}
\end{mathletters}
where
\begin{equation}
C\equiv \frac{4}{b}\left(\frac{\gamma_2}{\gamma_1(2-\epsilon)}\right)^2
\end{equation}

It is straightforward to see that, in the limits $\tau\rightarrow0$ and $\tau\rightarrow\infty$, the above expression reduce to the corresponding asymptotic expressions of Sections III-V.

\begin{tabular}{ccccccccll}
\multicolumn{10}{c}{TABLE 1 asymptotic solutions at early times: 
$\mid\!3+2\omega\!\mid \rightarrow(3/\lambda^2)(k+\mid\!\Phi-1\!\mid^{-\delta})$}\\ \hline\hline
\multicolumn{10}{c}{False Vacuum Inflationary models}\\ \hline\hline
$3+2\omega(0)$ & 
$y_0$ & 
\multicolumn{1}{c}{$a$} & 
\multicolumn{1}{c}{$\Phi$} & 
\multicolumn{1}{c}{$\xi$} &
\multicolumn{1}{c}{$a(0)$} &
$\Phi(0)$ &
\multicolumn{1}{c}{$\xi(0)$} & 
\multicolumn{2}{c}{Parameters} \\ \hline 

$>0$ & 		
$<0$ &		
$t^{\frac{1+\alpha}{1+3\alpha}}$ &
$t^{\frac{-2}{1+3\alpha}}$ & 
$ t^{-1}$ & 
$0$ &
$+\infty$ & 
$+\infty$ &
&
$k>0$\\

$>0$ & 		
$>0$ &		
$t^{\frac{\alpha-1}{3\alpha-1}}$ &
$t^{\frac{2}{3\alpha-1}}$ &
$ t^{-1}$ &
$\left\{\begin{array}{l}0\\+\infty \end{array}\right.$ &
$\begin{array}{l}0\\0\\0\end{array}$ & 
$\begin{array}{l}+\infty\\  -\infty\end{array}$ & 
$\begin{array}{l}\alpha>1\\ 2/3<\alpha<1 \end{array}$ &
$\begin{array}{l}k>-1\\ k>-1\end{array}$	\\ 

& 		
&		
&
&
const &
const &
$0$ & 
const & 
$\alpha=1$ &
$k>-1$	\\ 

$>0$ & 		
$=0$ &		
$t^{\frac{3\alpha^2-2}{2}}$ &
$t^2$ &
$t^{-1}$ &
$\left\{\begin{array}{l}0\\ +\infty \end{array}\right. $ &
$\begin{array}{l}0\\ 0 \end{array}$  & 
 $\begin{array}{l} +\infty\\ -\infty \end{array}$  & 
$\begin{array}{l} \alpha^2>2/3 \\ 
4/9<\alpha^2<2/3 \end{array}$ &
$\begin{array}{l} k>-1\\ k>-1\end{array}$
\\ 

 & 		
 &		
 &
 &
$t$ &
const &
$0$ & 
$0$ & 
$\alpha^2=2/3$&
$k>-1$
\\ \hline 

$=0^+$ & 
$<0$ & 
$t^{\frac{1}{1+\delta}}$ &
$t^{-\frac{2}{1+\delta}}$ &
$t^{-1}$ & 
$0$ &
$+\infty$ & 
$+\infty$ &
$\delta>0$ &
 $k=0$\\ 

$=0^-$ & 
$=0$ & 
$t^{\frac{1}{\delta-1}}$ &
$t^{-\frac{2}{\delta-1}}$ &
$t^{-1}$ & 
$0$ &
$+\infty$ & 
$+\infty$ &
$\delta>1$ &
 $k=0$\\ 

 & 
 & 
$t_{*}^{\frac{1}{\delta-1}}$ &
$t_{*}^{-\frac{2}{\delta-1}}$ &
$t_{*}^{-1}$ & 
$0$ &
$+\infty$ & 
0 &
$0<\delta<1$ &
 $k=0$\\ \hline 

$<0$ & 
$\neq0$ & 
const. &
const. &
$0$ & 
const &
const & 
$0$ &
&
\\

$<0$ & 
$=0$ & 
$t_{*}^{-\frac{3\alpha^2+2}{2}}$ &
$t_{*}^{2}$ &
$t_{*}^{-1}$ & 
$0$ &
$+\infty$ & 
$0$ &
$$ &
 $k>0$\\ \hline \hline
\multicolumn{10}{c}{Vacuum and Radiation-dominated models}\\ \hline\hline

$3+2\omega(0)$ & 
$y_0$ & 
\multicolumn{1}{c}{$a$} & 
\multicolumn{1}{c}{$\Phi$} & 
\multicolumn{1}{c}{$\xi$} &
\multicolumn{1}{c}{$a(0)$} & 
$\Phi(0)$ &
\multicolumn{1}{c}{$\xi(0)$} &
\multicolumn{2}{c}{Parameters} \\ \hline 

$>0$ & 	
$<0$&
$t^{\frac{1+\alpha}{3\alpha+1}}$ &
$t^{\frac{-2}{3\alpha+1}}$ & 
$ t^{\frac{1-\alpha}{1+3\alpha}}$ & 
$0$ &
+$\;\infty\;$ &
$\left\{ \begin{array}{l}+\infty \\ \mbox{const} \\0\end{array} \right.$ &
$\begin{array}{l} \alpha>1\\ \alpha=1\\ \alpha<1\end{array}$&
$\begin{array}{l} k>0\\ k>0\\ k>0\end{array}$\\

$>0$ & 		
$>0$ & 
$t^{\frac{\alpha-1}{3\alpha-1}}$ &
$t^{\frac{2}{3\alpha-1}}$ & 
$ t^{\frac{1+\alpha}{1-3\alpha}}$ &
$0$ &
$0$ &		
$\left\{ \begin{array}{l}+\infty \\ 0 \end{array} \right.$ &
$\begin{array}{l}\alpha>1 \\1/3<\alpha<1\end{array} $ &
$\begin{array}{l}k>-1\\ k>-1 \end{array}$ \\ \hline

$=0$ & 		
$<0$ & 
$t^{1/2}$ &
$t^{-1}$ & 
const. & 
$0$ &
+$\infty$ & 
$\mbox{const}$ &
 &
 $k=0$ \\

$=0$ & 		
$>0$ & 	
$t^{-1}$ &
$t^2$ & 
$-t^{-2}$ & 	
$>0$ &
$+\infty$	&
$-\infty$ &
 &
 $k=-1$ \\ \hline

$<0$ & 		
$>0$ &
const &
const &
const & 
$0$ &
$<1$ &	
 $-\mbox{const}$ &
 &
$k=-1$ \\

$<0$ & 		
$<0$ & 
const & 
const & 
const & 
$0$ &
$>1$ &
 $-\mbox{const}$ & 
$\lambda\leq 1/2$ &
$ k=-1$ \\ \hline \hline
\multicolumn{10}{c}{Matter-dominated models}\\ \hline\hline

$3+2\omega(0)$ & 
$y_0$ & 
\multicolumn{1}{c}{$a$} & 
\multicolumn{1}{c}{$\Phi$} & 
\multicolumn{1}{c}{$\xi$} &
\multicolumn{1}{c}{$a(0)$} & 
$\Phi(0)$ &
\multicolumn{1}{c}{$\xi(0)$} & 
\multicolumn{2}{c}{Parameters} \\ \hline 

$>0$ & 		
$<0$ &		
$t^{\frac{\alpha+1}{3\alpha+1}}$ &
$t^{\frac{-2}{3\alpha+1}}$ & 
$ t^{\frac{1}{2}\frac{1-3\alpha}{1+3\alpha}}$ & 
$0$ &
$\infty$ & 
$\left\{\begin{array}{l}0\\ \mbox{const}\\ +\infty \end{array}\right.$ &
$\begin{array}{l}\alpha<1/3\\ \alpha=1/3 \\ \alpha>1/3 \end{array}$ &
$k>0$\\

$>0$ & 		
$>0$ &		
$t^{\frac{\alpha-1}{3\alpha-1}}$ &
$t^{\frac{2}{3\alpha-1}}$ &
$ t^{\frac{1}{2}\frac{1+3\alpha}{1-3\alpha}}$ & 
$\left\{\begin{array}{l}0\\ +\infty \end{array}\right.$ &
$\begin{array}{l}0\\0\end{array}$ & 
$\begin{array}{l}+\infty\\ -\infty \end{array}$ & 
$\begin{array}{l}\alpha>1\\ 1/3<\alpha<1 \end{array}$ &
 $\begin{array}{l}k>-1\\k>-1\end{array}$	\\ 

& 		
&		
&
&
const &
const &
$0$ & 
const & 
$\alpha=1$ &
$k>-1$	\\ 

$>0$ & 		
$=0$ &		
$t^{2\frac{3\alpha^2-1}{9\alpha^2-1}}$ &
$t^{\frac{4}{9\alpha^2-1}}$ &
$t^{-\frac{2}{9\alpha^2-1}}$ &
$\left\{\begin{array}{l}0\\ +\infty \end{array}\right. $ &
$\begin{array}{l}0\\ 0 \end{array}$  & 
 $\begin{array}{l} +\infty\\ -\infty \end{array}$  & 
$\begin{array}{l} \alpha^2>1/3 \\ 
1/9<\alpha^2<1/3 \end{array}$ &
$\begin{array}{l} k>-1\\ k>-1\end{array}$
\\ 

 & 		
 &		
 &
 &
$t$ &
const &
$0$ & 
$0$ & 
$\alpha^2=1/3$&
$k>-1$
\\ \hline 

$=0^+$ & 
$<0$ & 
$t^{\frac{1}{1+\delta}}$ &
$t^{-\frac{2}{1+\delta}}$ &
$t^{-\frac{2\delta-1}{2(\delta+1)}}$ & 
$0$ &
$\infty$ & 
$\left\{\begin{array}{l}0\\ \mbox{const} \\+\infty \end{array}\right.$ &
$\begin{array}{l}\delta<1/2\\ \delta=1/2 \\ \delta>1/2\end{array}$ &
$\begin{array}{l} k=0\\ k=0\\ k=0\end{array}$\\ 

$=0^-$ & 
$=0$ & 
$t^{\frac{2}{2\delta+1}}$ &
$t^{-\frac{4}{2\delta+1}}$ &
$t^{-2\frac{\delta-1}{2\delta+1}}$ & 
$0$ &
$+\infty$ & 
$\left\{\begin{array}{l}0\\ \mbox{const} \\+\infty\end{array}\right.$ &
$\begin{array}{l}0<\delta<1\\ \delta=1 \\ \delta>1\end{array}$ &
 $k=0$\\ \hline

$<0$ & 
$\neq0$ & 
const. &
const. &
$0$ & 
$>0$ &
$\mbox{const}$ & 
$0$ &
&
\\

$<0$ & 
$=0$ & 
$t^{2\frac{1+3\alpha^2}{1+9\alpha^2}}$ &
$t^{-\frac{4}{1+9\alpha^2}}$ &
$t^{\frac{2}{1+9\alpha^2}}$ & 
$0$ &
$+\infty$ & 
$0$ &
$$ &
 $k>0$\\ \hline \hline
{\footnotesize NOTES:} & \multicolumn{9}{l}{\footnotesize
1) Vacuum ($\Gamma=0$) models require $3+2\omega(0)\geq 0^+$.}\\
 & \multicolumn{9}{l}{\footnotesize
2) $\alpha=\sqrt{k}/\lambda$ when $\Phi(0)=\infty$, and $\alpha=\sqrt{k+1}/\lambda$ when $\Phi(0)=0$}\\
& \multicolumn{9}{l}{\footnotesize
3) $t_{*}$ denotes that early-time solutions correspond to the limit $t\rightarrow-\infty$}
\end{tabular}

\newpage

\begin{figure}
\centerline{\epsfig{figure=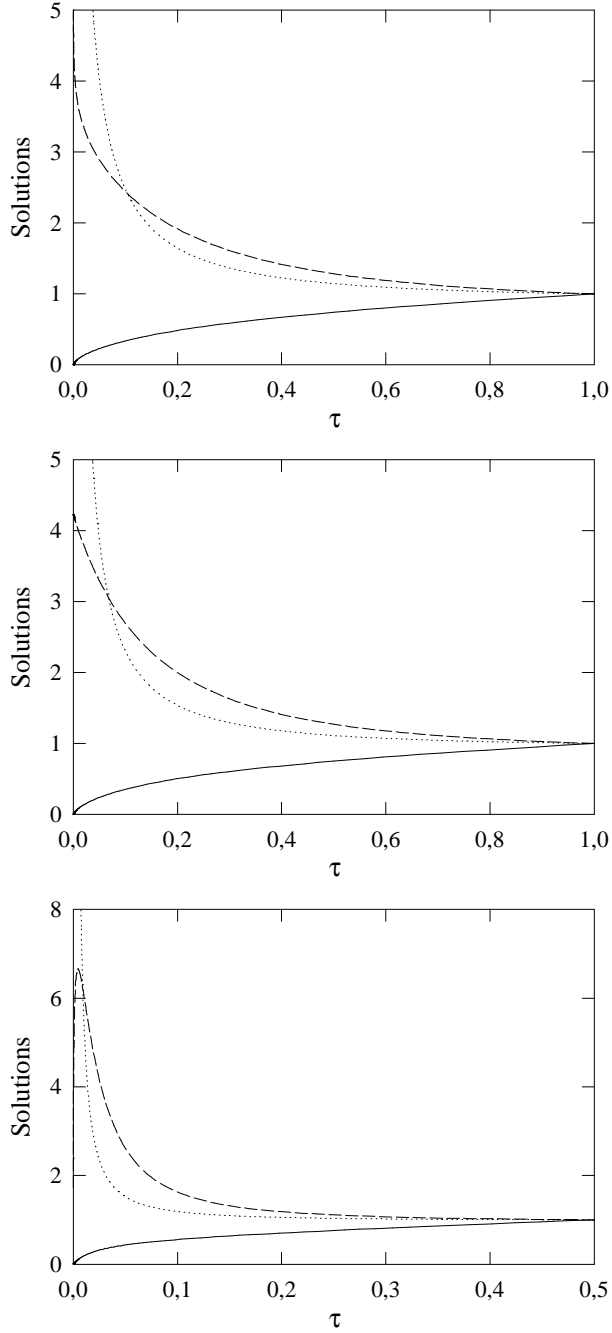,width=8cm}}
\vspace{0.5cm}
\caption{Models with $W>0$, $\alpha\neq 0$ and $y_0<0$: numerical solutions obtained by considering the theory defined, at any time, by $W(\tau)=(p+q\tau)^3$ with $b=1$, $\gamma=1$, and a) $\alpha^2=0.2$, $y_0=-3$,
b) $\alpha^2=\alpha_{c}^{2}=1/9$, $y_0=-3$, c) $\alpha^2=0.01$, $y_0=-1$. Solid lines represent the scale factor, dotted lines represent the scalar field, and
dashed lines represent the speed-up factor.}
\end{figure}

\begin{figure}
\centerline{\epsfig{figure=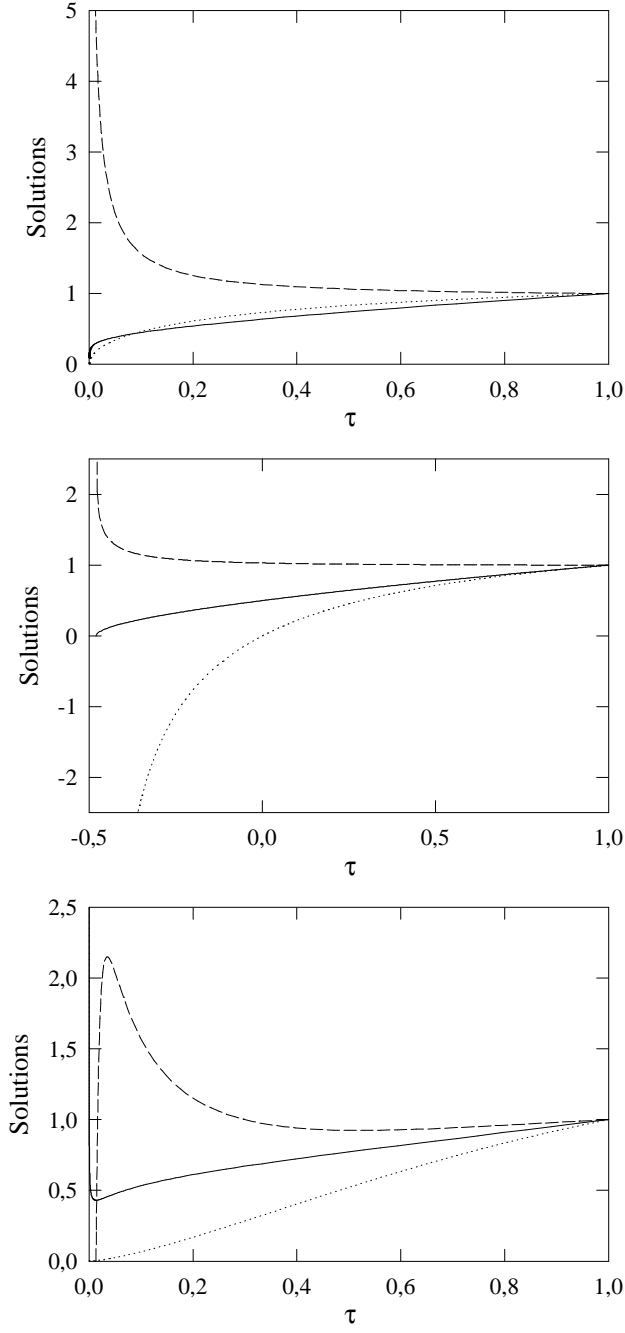,width=8cm}}
\vspace{0.5cm}
\caption{Models with $W>0$, $\alpha\neq 0$ and $y_0>0$: matter-dominated ($\gamma=1$)  solutions obtained 
by considering the theory defined, at any time, by a) $W(\tau)=\alpha^2+ b\tau^{1/2}$ with $\alpha^2=25/9$, $b=1$, $y_0=1$,
b) $W(\tau)=\alpha^2+ b\tau$ with $\alpha^2=\alpha_{c}^{2}=1$, $b=1$, $y_0=1$, c) $\alpha^2=4/9$, $b=0.5$, $y_0=0.02$. Symbols for lines are the same as in Figure 1}
\end{figure}

\begin{figure}
\centerline{\epsfig{figure=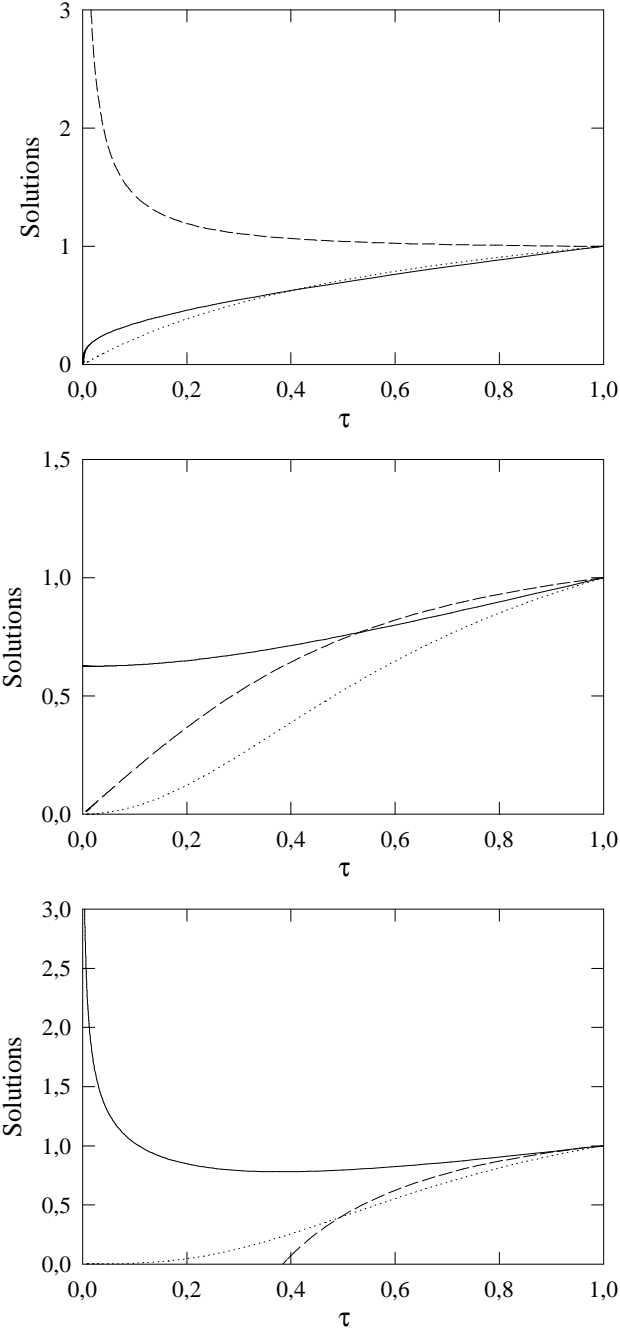,width=8cm}}
\vspace{0.5cm}
\caption{Models with $\alpha\neq 0$ and $y_0=0$: matter-dominated solutions 
obtained by considering the theory defined, at any time, by $b=1$ and
a) $W(\tau)=\alpha^2+b\tau$  ($\alpha^2=5/9$),  
b) $W(\tau)=\alpha^2+b\tau^2$ ($\alpha^2=\alpha_{c}^{2}=1/3)$, 
c) $W(\tau)=(\alpha+q\tau^3)^2$ ($b=2\alpha q$ and $\alpha^2=7/27$). Symbols for lines are the same as in Figure 1}
\end{figure}

\begin{figure}
\centerline{\epsfig{figure=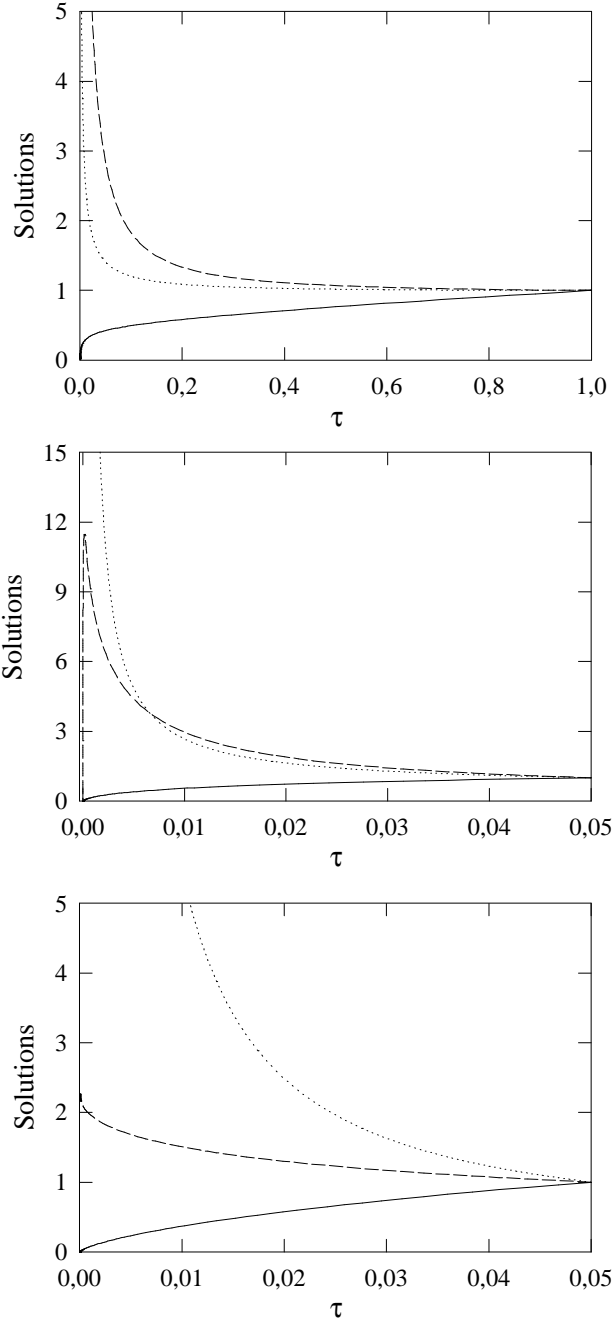,width=8cm}}
\vspace{0.5cm}
\caption{Models with $\alpha= 0$ and $W>0$: numerical solutions 
obtained by considering the theory defined, at any time, by 
$W(\tau)=b\tau(1+c\tau^{1/2})^3$ ($b, c>0$) with $y_0=-2$ and 
a) $\gamma=1$ ($\delta>\delta_c=1/2$), $b=0.1$, and $c=10$, 
b) $\gamma=1.12$ ($\delta<\delta_c=0.68$, $b=0.1$, and $c=10$,
c) $\gamma=10/9$ ($\delta=\delta_c=2/3$), $b=1$, and $c=0.5$.
Symbols for lines are the same as in Figure 1}
\end{figure}

\begin{figure}
\centerline{\epsfig{figure=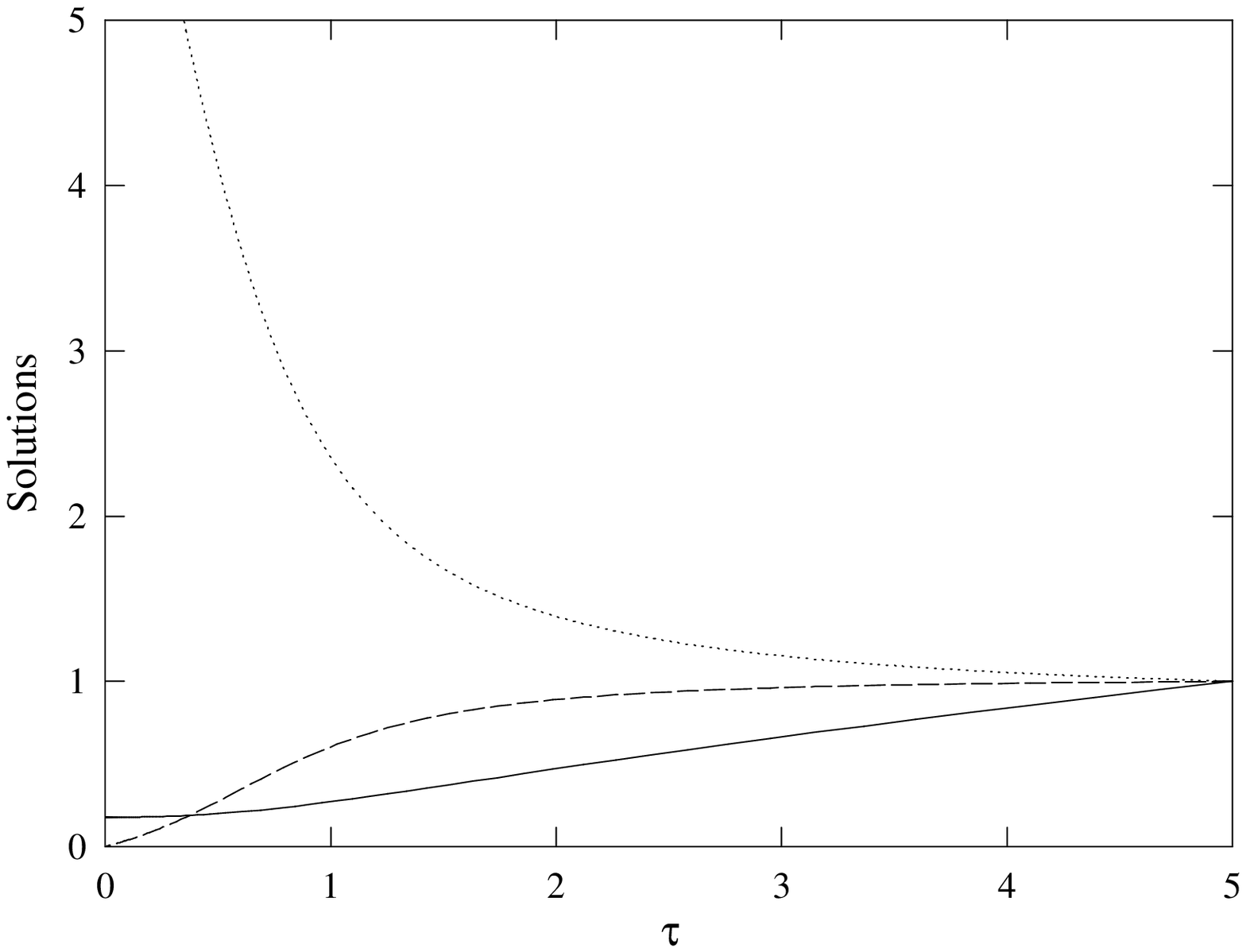}}
\vspace{0.5cm}
\caption{Models with $W<0$, $\alpha\neq 0$ and $y_0<0$: numerical solutions obtained by 
considering the theory defined, at any time, by $W(\tau)=\alpha^2+b\tau$  with $\gamma=1$, $\alpha^2=1$, $b=1$, $y_0=-2$.
Symbols for lines are the same as in Figure 1}
\end{figure}

\begin{figure}
\centerline{\epsfig{figure=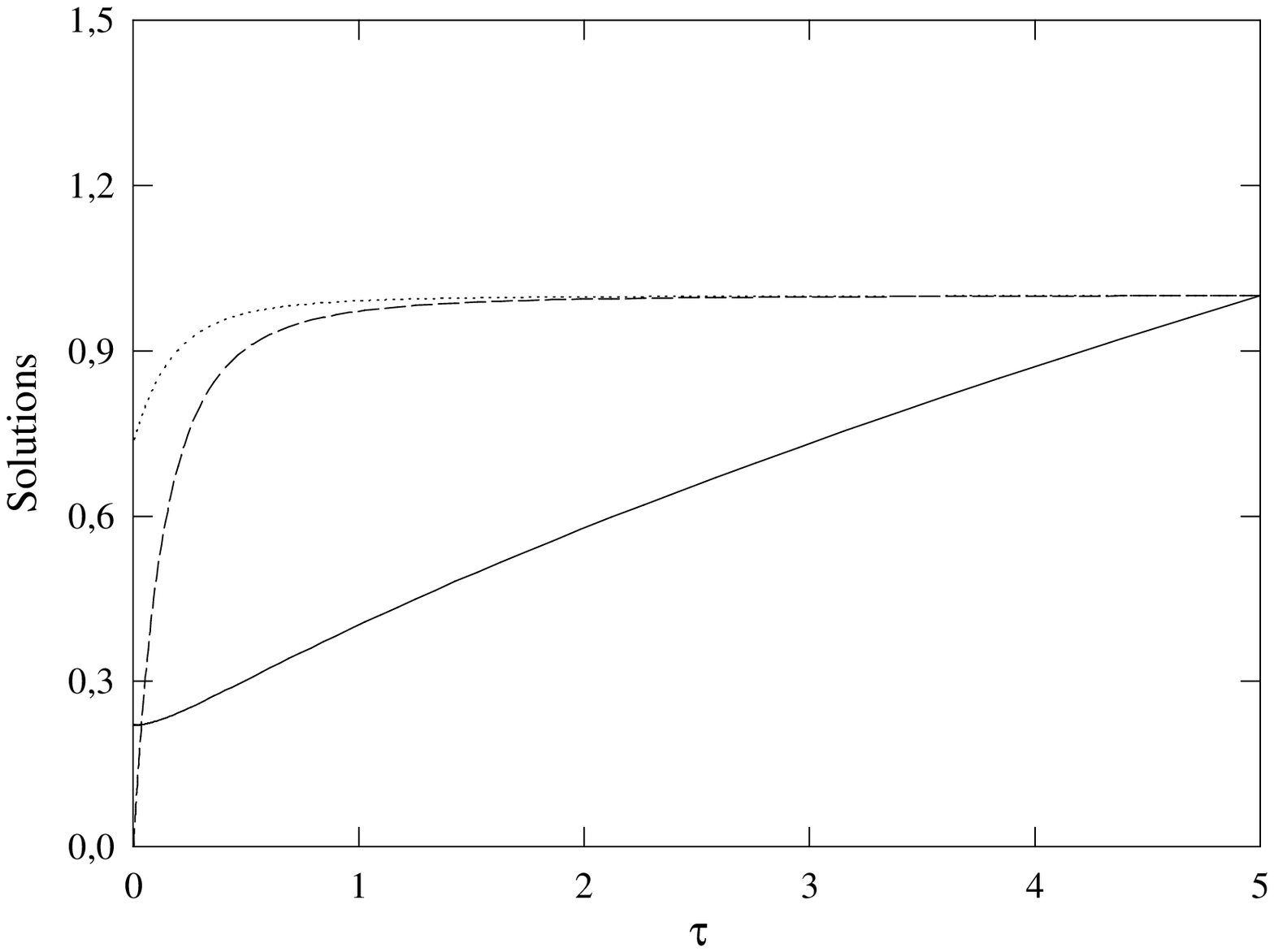}}
\vspace{0.5cm}
\caption{Models with $W<0$, $\alpha\neq 0$ and $y_0>0$: numerical solutions obtained by 
considering the theory defined, at any time, by $W(\tau)=(p+q\tau)^3$ with $\gamma=1$, $\alpha^2=1$, $b=10$, $y_0=1$.
Symbols for lines are the same as in Figure 1}
\end{figure}

\begin{figure}
\centerline{\epsfig{figure=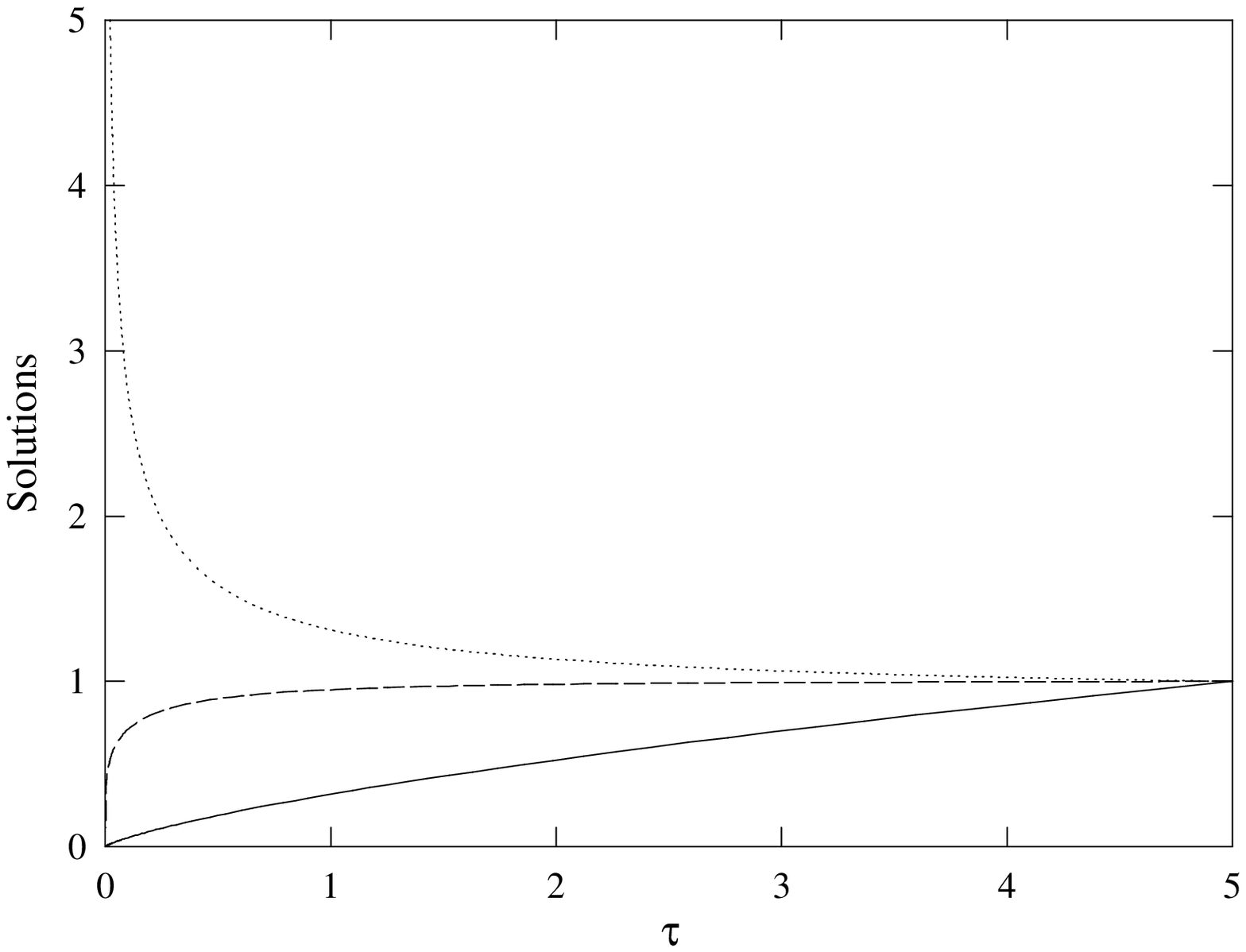}}
\vspace{0.5cm}
\caption{Models with $W<0$, $\alpha\neq 0$ and $y_0=0$: numerical solutions obtained by 
considering the theory defined, at any time, by $|W(\tau)|=\alpha^2+b\tau$
with $\gamma=1$, $\alpha^2=1$, $b=1$. Symbols for lines are the same as in Figure 1}
\end{figure}

\begin{figure}
\centerline{\epsfig{figure=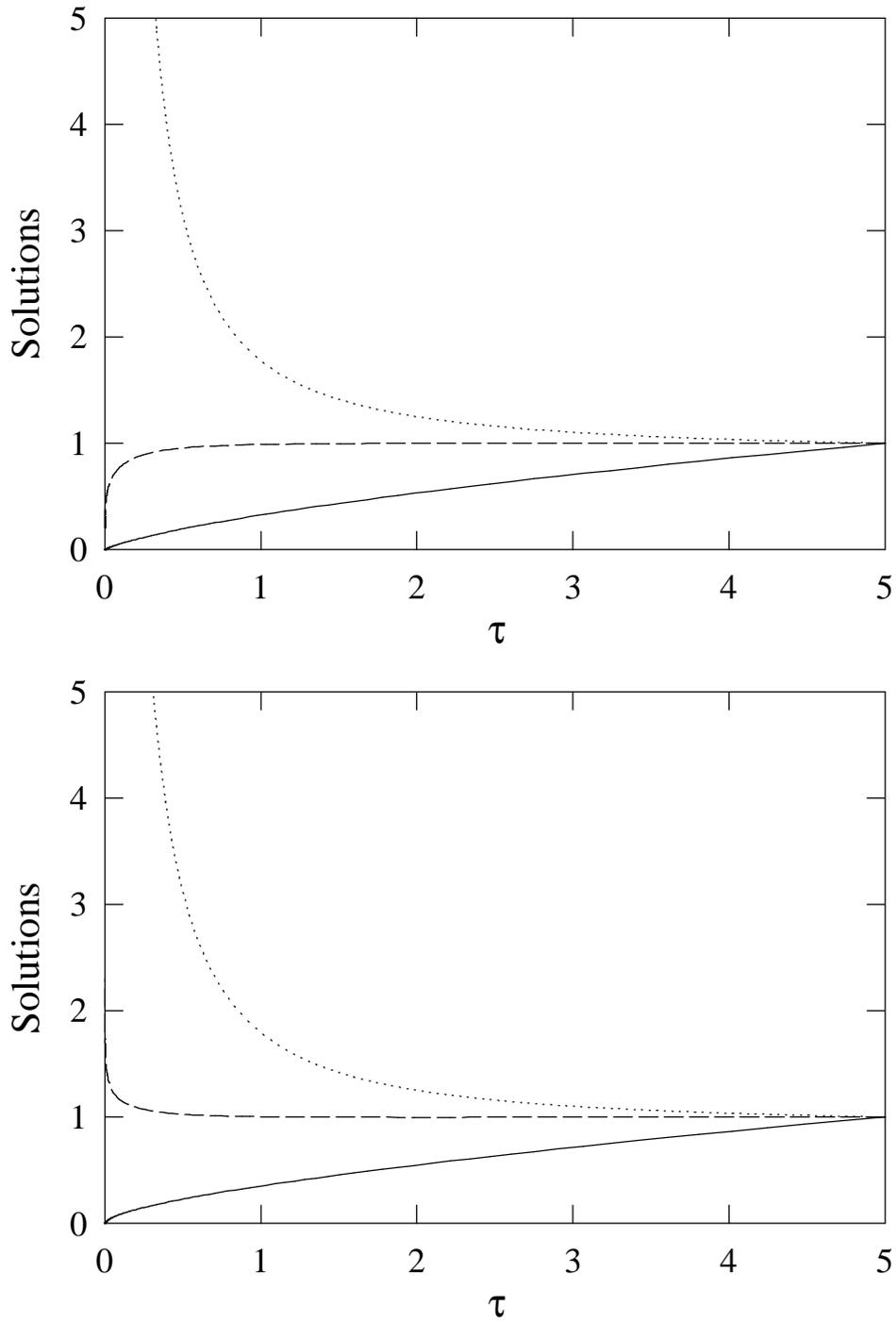}}
\vspace{0.5cm}
\caption{Models with $W<0$ and $\alpha=0$: analytical solutions obtained by 
considering the theory defined, at any time, by $|W(\tau)|=b\tau^\epsilon$ 
with $b=1$, $\gamma=1$ (so that the critical value $\delta_c=1$ corresponds to $\epsilon_c=4/3$) and a) $\epsilon=1.2$,  
b) $\epsilon=1.4$. Symbols for lines are the same as in Figure 1}
\end{figure}

\end{document}